\documentclass[journal,hidelinks]{IEEEtran}

\ifCLASSINFOpdf
\else
\usepackage[dvips]{graphicx}
\fi
\usepackage{url}
\usepackage{dblfloatfix}

\usepackage[dvipsnames]{xcolor}

\usepackage{graphicx} 
\usepackage{booktabs} 
\usepackage{mathrsfs}
\usepackage{amsmath}
\usepackage{amssymb}
\usepackage{bm}
\usepackage{mathtools}
\usepackage{cite}
\usepackage[acronym,shortcuts]{glossaries}
\usepackage{enumitem} 
\usepackage{comment}
\usepackage{algorithm, algpseudocode}
\usepackage{mathtools}
\usepackage{lipsum}
\usepackage{mleftright}
\usepackage{orcidlink}
\usepackage{soul}       
\usepackage{adjustbox}
\usepackage{threeparttable}
\usepackage{array}
\usepackage{amsthm}  
\setlist[itemize]{noitemsep} 
\usepackage{subfigure}

\providecommand{\UL}{\mathrel{\raise-4pt\hbox{\hglue -2.8ex
\vrule height .1ex width 2.3ex
\vrule height 3ex width .1ex
\hglue .4ex}}}
\providecommand{\ul}{\mathrel{\raise-2pt\hbox{\hglue -2.3ex
\vrule height .1ex width 2ex
\vrule height 2ex width .1ex
\hglue .4ex}}}
\providecommand{\ub}{
\mathrel{\hbox{\hglue -2.8ex \vrule height 2ex width .06ex}
\raise2ex\hbox{\hglue -0.1ex\vrule height .1ex width 2.5ex}
\hbox{\hglue -0.1ex \vrule height 2ex width .1ex
\hglue .4ex}}}
\providecommand{\lb}{
\mathrel{\raise-0.5ex
\hbox{\hglue -2.8ex \vrule height 2ex width .1ex
\vrule height .1ex width 2.5ex
\vrule height 2ex width .1ex
\hglue .4ex}}}
\providecommand{\UB}{
\mathrel{
\raise-1ex\hbox{\hglue -1.8em \vrule height 3ex width .1ex}
\raise2ex\hbox{\hglue -0.1ex\vrule height .1ex width 1.5em}
\raise-1ex\hbox{\hglue -0.1ex \vrule height 3ex width .1ex
\hglue .4ex}}}
\providecommand{\LB}{
\mathrel{\raise-1ex
\hbox{\hglue -1.8em \vrule height 3ex width .1ex
\vrule height .1ex width 1.5em
\vrule height 3ex width .1ex
\hglue .4ex}}}
\usepackage{trimclip}
\newif\iflclip
\newif\ifbclip
\newif\ifrclip
\newif\iftclip
\def\CLIP{\dimexpr\fboxrule+.2pt\relax}
\def\nulclip{0pt}
\newcommand\partbox[2]{%
\lclipfalse\bclipfalse\rclipfalse\tclipfalse%
\let\lkern\relax\let\rkern\relax%
\let\lclip\nulclip\let\bclip\nulclip\let\rclip\nulclip\let\tclip\nulclip%
\parseclip#1\relax\relax%
\iflclip\def\lkern{\kern\CLIP}\def\lclip{\CLIP}\fi
\ifbclip\def\bclip{\CLIP}\fi
\ifrclip\def\rkern{\kern\CLIP}\def\rclip{\CLIP}\fi
\iftclip\def\tclip{\CLIP}\fi
\lkern\clipbox{\lclip{} \bclip{} \rclip{} \tclip}{\fbox{#2}}\rkern%
}
\def\parseclip#1#2\relax{%
\ifx l#1\lcliptrue\else
\ifx b#1\bcliptrue\else
\ifx r#1\rcliptrue\else
\ifx t#1\tcliptrue\else
\fi\fi\fi\fi
\ifx\relax#2\relax\else\parseclip#2\relax\fi
}
\usepackage{efbox}

\newcommand\Tstrut{\rule{0pt}{2.6ex}}         

\DeclareMathOperator{\acos}{acos}

\newacronym{IoV}{IoV}{internet of vehicles}
\newacronym{SVD}{SVD}{singular value decomposition}
\newacronym{DCM}{DCM}{double-centering matrix}
\newacronym{3D}{3D}{three-dimensional}
\newacronym{GA}{GA}{genie-aided}
\newacronym{EA}{EA}{``\emph{estimate-then-average}''}
\newacronym{AE}{AE}{``\emph{average-then-estimate}''}
\newacronym{IRS}{IRS}{intelligent reflecting surface}
\newacronym{RSSI}{RSSI}{received signal strength indicator}
\newacronym{SotA}{SotA}{state-of-the-art}
\newacronym{CSI}{CSI}{channel state information}
\newacronym{D2D}{D2D}{device-to-device}
\newacronym{RR}{RR}{round-robin}
\newacronym{DA}{DA}{Dutch auction}
\newacronym{AV}{AV}{autonomous vehicle}
\newacronym{CWFL}{CWFL}{clustered WFL}
\newacronym{WFL}{WFL}{wireless federated learning}
\newacronym{RSMA}{RSMA}{rate splitting multiple access}
\newacronym{IoT}{IoT}{Internet-of-Things}
\newacronym{TDMA}{TDMA}{time-domain multiple access}
\newacronym{NOMA}{NOMA}{non-orthogonal multiple access}
\newacronym{ML}{ML}{machine learning}
\newacronym{MIMO}{MIMO}{multiple-input multiple-output}
\newacronym{CT}{CT}{compute-then-transmit}
\newacronym{FP}{FP}{fractional programming}
\newacronym{CF-mMIMO}{CF-mMIMO}{cell free massive MIMO}
\newacronym{iid}{i.i.d.}{independent and identically distributed}
\newacronym{AD}{AD}{autonomous driving}
\newacronym{DL}{DL}{downlink}
\newacronym{UL}{UL}{uplink}
\newacronym{IC}{IC}{interference cancellation}
\newacronym{SIC}{SIC}{successive interference cancellation}
\newacronym{BS}{BS}{base station}
\newacronym{TX}{TX}{transmit}
\newacronym{RX}{RX}{receive}
\newacronym{MU}{MU}{multi-user}
\newacronym{SISO}{SISO}{single-input single-output}
\newacronym{AWGN}{AWGN}{additive white Gaussian noise}
\newacronym{SINR}{SINR}{signal-to-interference-and-noise ratio}
\newacronym{FL}{FL}{federated learning}
\newacronym{CPU}{CPU}{central processing unit}
\newacronym{KNN}{KNN}{K-nearest-neighbor}
\newacronym{RF}{RF}{radio frequency}
\newacronym{GD}{GD}{gradient descent}
\newacronym{V2X}{V2X}{vehicle-to-anything}

\newacronym{RSS}{RSS}{received signal strength}
\newacronym{FIM}{FIM}{Fisher information matrix}
\newacronym{ToA}{ToA}{time of arrival}
\newacronym{ToF}{ToF}{time of flightl}
\newacronym{AoA}{AoA}{angle of arrival}
\newacronym{ADoA}{ADoA}{angle difference of arrival}
\newacronym{GP}{GP}{Gaussian process}
\newacronym{2D}{2D}{two-dimensional}
\newacronym{GPR}{GPR}{Gaussian process regression}
\newacronym{GNSS}{GNSS}{global navigation satellite systems}
\newacronym{B5G}{B5G}{beyond fifth-generation}
\newacronym{6G}{6G}{sixth-generation}
\newacronym{RRH}{RRH}{remote radio head}
\newacronym{GPS}{GPS}{Global Positioning System}
\newacronym{RFID}{RFID}{radio frequency identification}
\newacronym{TCAS}{TCAS}{traffic alert and collision avoidance systems}
\newacronym{RMSE}{RMSE}{root mean square error}
\newacronym{MSE}{MSE}{mean square error}
\newacronym{SGD}{SGD}{stochastic gradient descent}
\newacronym{PDF}{PDF}{probability density function}
\newacronym{CU}{CU}{computing unit}
\newacronym{DM-MIMO}{DM-MIMO}{distributed massive multiple-input multiple-output}
\newacronym{LOS}{LOS}{line-of-sight}
\newacronym{NLOS}{NLOS}{non-line-of-sight}
\newacronym{ROI}{ROI}{region of interest}
\newacronym{AP}{AP}{access point}
\newacronym{TDOA}{TDOA}{time difference of arrival}
\newacronym{UE}{UE}{user equipment}
\newacronym{dB}{dB}{decibel}
\newacronym{RIS}{RIS}{reconfigurable intelligent surface}
\newacronym{CG}{CG}{conjugate gradient}

\newacronym{PG}{PG}{proximal gradient}
\newacronym{SVT}{SVT}{singular value thresholding}
\newacronym{NN}{NN}{nuclear norm}
\newacronym{NMSE}{NMSE}{normalized mean square error}
\newacronym{MC}{MC}{matrix completion}
\newacronym{NP}{NP}{non-deterministic polynomial-time}
\newacronym{EDM}{EDM}{euclidean distance matrix}
\newacronym{SC}{SC}{soft-connected}
\newacronym{CRLB}{CRLB}{Cram{\'e}r-Rao Lower Bound}
\newacronym{PoA}{PoA}{phase of arrival}
\newacronym{UAV}{UAV}{unmanned aerial vehicle}
\newacronym{VR}{VR}{virtual reality}
\newacronym{MDS}{MDS}{multidimensional scaling}
\newacronym{SMDS}{SMDS}{super multidimensional scaling}

\newacronym{RBL}{RBL}{rigid body localization}
\newacronym{RBT}{RBT}{rigid body tracking}
\newacronym{SC-RBL}{SC-RBL}{soft-connected RBL}
\newacronym{W-RBL}{W-RBL}{\underline{wireless} RBL}

\newacronym{SDP}{SDP}{semidefinite programming}
\newacronym{ISAC}{ISAC}{integrated sensing and communications}
\newacronym{SDR}{SDR}{semi-definite relaxation}

\newacronym{OPP}{OPP}{orthogonal Procrustes problem}
\newacronym{SLAM}{SLAM}{simultaneous localization and mapping}
\newacronym{WLS}{WLS}{weighted least square}
\newacronym{SI}{SI}{soft-impute}

\newacronym{CSDP}{CSDP}{constrained semidefinite programming}
\newacronym{iff}{iff}{if and only if}
\newacronym{UDPN}{UDPN}{unit disk planar network}
\newacronym{BFS}{BFS}{Breadth First Search}
\newacronym{AODV}{AODV}{Ad hoc On-Demand Distance Vector}
\newacronym{GaBP}{GaBP}{Gaussian Belief Propagation}

\newacronym{LEO}{LEO}{low earth orbit}
\newacronym{PEB}{PEB}{position error bound}

\newtheorem{lemma}{Lemma}

\begin{document}

\title{Fundamental Limits of Rigid Body Localization\vspace{-1ex}}

\author{\IEEEauthorblockN{Niclas~F\"uhrling\textsuperscript{\orcidlink{0000-0003-1942-8691}},~\IEEEmembership{Graduate Student Member,~IEEE,} Ivan Alexander Morales Sandoval\textsuperscript{\orcidlink{0000-0002-8601-5451}},~\IEEEmembership{Graduate Student Member,~IEEE,} Giuseppe Thadeu Freitas de Abreu\textsuperscript{\orcidlink{0000-0002-5018-8174}},~\IEEEmembership{Senior Member,~IEEE,} Gonzalo Seco-Granados\textsuperscript{\orcidlink{0000-0003-2494-6872}},~\IEEEmembership{Fellow,~IEEE,} David~Gonz{\'a}lez~G.\textsuperscript{\orcidlink{0000-0003-2090-8481}},~\IEEEmembership{Senior Member,~IEEE,\vspace{-4ex}} and Osvaldo~Gonsa\textsuperscript{\orcidlink{0000-0001-5452-8159}}}\vspace{-1.5ex}

\thanks{N.~F\"uhrling, I.~A.~Morales Sandoval and G.~T.~F.~de~Abreu are with the School of Computer Science and Engineering, Constructor University, Campus Ring 1, 28759, Bremen, Germany {(e-mails: [nfuehrling, imorales, gabreu]@constructor.university).}}
\thanks{G.~Seco-Granados is with the Department of Telecommunications and Systems Engineering, Universitat Autònoma de Barcelona, Spain (email: gonzalo.seco@uab.cat).}
\thanks{D.~Gonz{\'a}lez~G. and O.~Gonsa are with {\color{black}AUMOVIO Germany GmbH}, Guerickestrasse 7, 60488, Frankfurt am Main, Germany (e-mail: david.gonzalez.g@ieee.org, {\color{black}osvaldo.gonsa@aumovio.com}
).}
}

\setlength{\parskip}{0pt}

\markboth{To be submitted to the IEEE for possible publication}
{Shell \MakeLowercase{\textit{et al.}}: Bare Demo of IEEEtran.cls for IEEE Journals}
\maketitle

\begin{abstract}

We consider a novel {\color{black}and general approach to easily compute the \acp{CRLB} of \ac{RBL} problem using arbitrary types of information}.
To that end, we adopt an information-centric construction of the \ac{FIM}, which allows {\color{black}capturing} the contribution of each measurement towards the \ac{FIM} {\color{black}explicitly}, both in terms of input measurement types, as well as of their error distributions.
Taking advantage of this approach, we derive a generic framework for {\color{black}evaluating the} \ac{CRLB}, which is applicable to {\color{black}arbitrary} rigid body localization scenarios, {\color{black}and which, unlike the formulation of FIM commonly used in point-target localization, is better suited to RBL problems as it explicitly allows capturing the precision in} both the translation vector and the rotation matrix (or alternative the rotation angles) {\color{black}of the rigid body}, with respect to a reference.
{\color{black} Examples of \acp{CRLB} obtained via the proposed approach are given in closed form}, including the bound incorporating an orthonormality constraint onto the rotation matrix{\color{black}, which enables a straightforward adjustment of the derived bound when new measurements are added or removed.}
Numerical results illustrate that the derived expression correctly lower-bounds the errors of estimated localization parameters obtained via various related \ac{SotA} estimators, revealing their accuracies and suggesting that \ac{SotA} \ac{RBL} algorithms can still be improved.
\end{abstract}

\vspace{-0.5ex}

\IEEEpeerreviewmaketitle

\glsresetall

\vspace{-2ex}
\section{Introduction}
\vspace{-1ex}
\IEEEPARstart{L}{ocalization} algorithms are of great interest in wireless system applications such as autonomous driving \cite{Whiton_22}, robotics \cite{Correll_22}, augmented reality \cite{Baker_24}, \ac{IoV} \cite{WangTIV2020} and digital twins \cite{ManickamAcces2023}, which implies that understanding the limits of their performance is of essential importance to determining the feasibility of such applications.
The most prominent of these bounds is the \ac{CRLB}, which provides a fundamental limit on the variance of unbiased estimates of a given parameter \cite{Rao_92}, applicable not only in the context of localization \cite{Gezici2008}, but also in many other fields, such as machine learning \cite{Bader2024}, radar systems \cite{Rayan_2024}, and communications \cite{Meng_2024}.
The \ac{CRLB} is derived from the \ac{FIM}, which captures the amount of information that a set of measurements provides about the parameters of interest. 
Although other well known bounds such as the Bhattacharyya bound \cite{Bhattacharyya_1947}, the Hodges-Lehmann bound \cite{Fraser1952} or the Weiss-Weinstein bound \cite{Weiss1980} also exist, the \ac{CRLB} is the most widely used in the literature, as it is relatively easy to compute compared to more complex bounds.

Wireless localization \cite{burghal_2020, obeidat2021review, ShanAccess2023} can be seen as an early example of \ac{ISAC}, in so far as it demonstrates how communication signals can be repurposed to acquire situational awareness, which is increasingly seen as a key added value of \ac{B5G} \cite{WangJSAC2022} and \ac{6G} \cite{02:00074} systems, as recent advancements in \ac{ISAC} technology \cite{Zhang_2021,Rou_2024} have indeed shown that radar parameters ($i.e.$, range, bearing and velocity) can be extracted from conventional communications signals \cite{Rayan_2024,Rayan_Journal}.
This can be done both actively, $i.e.$, via signals transmitted by the target, or passively, $i.e.$, via round-trip reflections of signals transmitted by the sensors themselves, resulting in richer and more abundant positioning information.

As a result of the above, there is growing interest in the \acf{RBL} problem \cite{WangTSP2020, Nic_RBL_WP, FuehrlingV2X2024}, which extends beyond single-point localization towards systems capable of inferring not only the location of point targets, but also the shape and orientation of extended targets represented by a set of pre-defined landmark points.
However, in spite of the vast variety of approaches and \ac{SotA} algorithms to solve the \ac{RBL} problem \cite{YuITJ2023, ZhaISEEIE2021, DongTWC2023, ChepuriTSP2014, Zhou_2019}, a generic bound on the accuracy parameter estimates applicable to any type of \ac{RBL} scenario has not yet appeared, which is a key requirement for the design of robust localization systems and the corresponding evaluation of their performance.

In order to obtain the \ac{CRLB} for any estimation problem, a corresponding \ac{FIM} needs to be constructed, which is based on the \ac{PDF} of the associated measurements, calculated by taking the expected value of the negative second derivative of the log-likelihood function with respect to the parameter vector that captures the amount of information that the observed data carries about the parameters.
This is a well known and widely used approach in the literature, which will be he hereafter referred to as the \textit{element-centric} \ac{FIM} formulation \cite{Rao_92}.

{\color{black} 
In light of this approach, many works have derived \acp{CRLB} for specific localization applications.
For instance, \ac{LEO}-based localization was considered in \cite{Emenonye_2025}, while holographic positioning was addressed in \cite{Damico_2022}.
Bounds under other scenarios and for various applications have also been derived, $e.g.$, in \cite{Chen_2021, Angjelichinoski_2015, He_2024, ShenTIT2010Part1, ShenTIT2010Part2}.
These works provide valuable insights into the performance limits of specific localization scenarios, however, they do not offer a general framework for deriving the \ac{CRLB} for the \ac{RBL} problem specifically.

The key distinction between \ac{RBL} and the more conventional and widely studied point-target localization is that in \ac{RBL} the landmark points defining a given \ac{3D} object are rigidly ``locked'' together, such that instead of the location of the landmark points themselves, one is interested in estimating the translation and rotation parameters of the rigid body, relative to a known prior reference position.

It follows that bounding the errors of \ac{RBL} algorithms via the \ac{CRLB} approach requires the evaluation of \acp{FIM} that explicitly capture the precision of these parameters, rather than the location of the landmark points individually.
This requirement when deriving \ac{RBL}-specific \ac{CRLB} has been partially addressed in related literature, but only for certain scenarios and under limited types or observations and error models.
For instance, a unitarily constrained \ac{CRLB} for the 3D \ac{RBL} case was derived in \cite{ChepuriTSP2014}, but for the range-based estimation case only, and under the assumption that measurement errors are Gaussian.
In turn, in \cite{Chen_2015}, unconstrained \acp{CRLB} with generic measurement covariances were derived for stationary and moving 2D \ac{RBL} problems, while the constrained 3D case was only briefly discussed, without being fully addressed.
Besides the aforementioned limitations, the \acp{CRLB} given in \cite{ChepuriTSP2014} and  \cite{Chen_2015} have in common the standard element-centric approach to constructing the \ac{FIM}, whereby each of its entry is obtained with basis on the entire likelihood function of all observations under the whole set of desired estimate parameters, which can be very cumbersome to evaluate, especially when dealing with complex scenarios involving multiple types of measurements and error distributions.

To the best of our knowledge, no general and comprehensive approach to compute \acp{CRLB} for \ac{RBL} problems have yet been given, which is the goal of this article.
In so doing, we shall also build on a modern approach to constructing \acs{FIM} \cite{Ghods_2014}, whereby each input measurement, as well as the corresponding type of error, is accounted for independently\footnote{\color{black}This assumes, of course, that measurements are obtained independently, which however is the case in the vast majority of real-life applications, and which in any case yields an absolute limit.}}.
Such an approach, here referred to as the \textit{information-centric} \ac{FIM} formulation, defines the \ac{FIM} in a sum-product form that can be applied and extended to any type of localization scenario, including the \ac{RBL} problem.

All in all, we propose in this article a novel formulation of the \ac{CRLB} suitable to \ac{RBL}, namely, aimed at determining the fundamental limit on the estimate of the translation and the rotation of a rigid body relative to a given reference {\color{black} position}.
To that extend, we first revisit the original information-centric \ac{FIM} formulation for a target group of independent nodes {\color{black}\cite{Ghods_2014}}, which is then extended to the case of a rigid body, evaluating it with respect to the rotation and translation parameters by making use of the simple sum-product formulation of the \ac{FIM}.
The proposed {\color{black}\ac{CRLB} expression} can be applied to any type of rigid body localization scenario, where we additionally provide the closed-form expressions for specific case of the constrained bound for the rotation matrix, since the rotation matrix is part of the special orthogonal group.

The contributions of this article are summarized as follows:
\vspace{-3ex}
\begin{itemize}
{\color{black}\item Leveraging an information-centric approach to constructing the \ac{FIM}   \cite{Ghods_2014}, a novel general framework to obtain \acp{CRLB} for \ac{RBL} is derived, which captures the fundamental limits of the estimates of translation and rotation angles of rigid bodies, relative to a reference position.
\item Closed-form expressions for the derived \acp{CRLB} are offered for multiple measurement types and error distributions, which can be directly applied to any \ac{RBL} scenario and enable a straightforward calculation of the bounds if new measurements are added or removed.
\item Addressing the orientation estimation further, a variation of the latter bound is also derived for the case where the entire rotation matrix, rather than individual angles, is obtained, where the rotation matrix is constrained to be in the special orthogonal group.
}
\end{itemize}

The remainder of the article is {\color{black}organized} as follows.
First, a description of the rigid body representation and a description of the information-centric \ac{FIM} formulation {\color{black}are} offered in Section \ref{sec:prior}.
Then, in Section \ref{sec:prop}, the proposed derivation of the \acp{CRLB} for the parameters estimated in \ac{RBL} problems is {\color{black}explained}.
Finally, a numerical evaluation of the proposed bounds is presented in Section \ref{sec:res}, along with comparisons with various related \ac{SotA} \ac{RBL} algorithms.
Conclusions and some words on possible future work then follow in Section \ref{sec:conclusions}, and some additional material aimed at facilitating understanding of some the work are offered in Appendices.

\section{Preliminaries}
\label{sec:prior}

\subsection{Rigid Body Representation Model}
\label{sec:RBL}

\begin{figure}[H]
\centering
\includegraphics[width=\columnwidth]{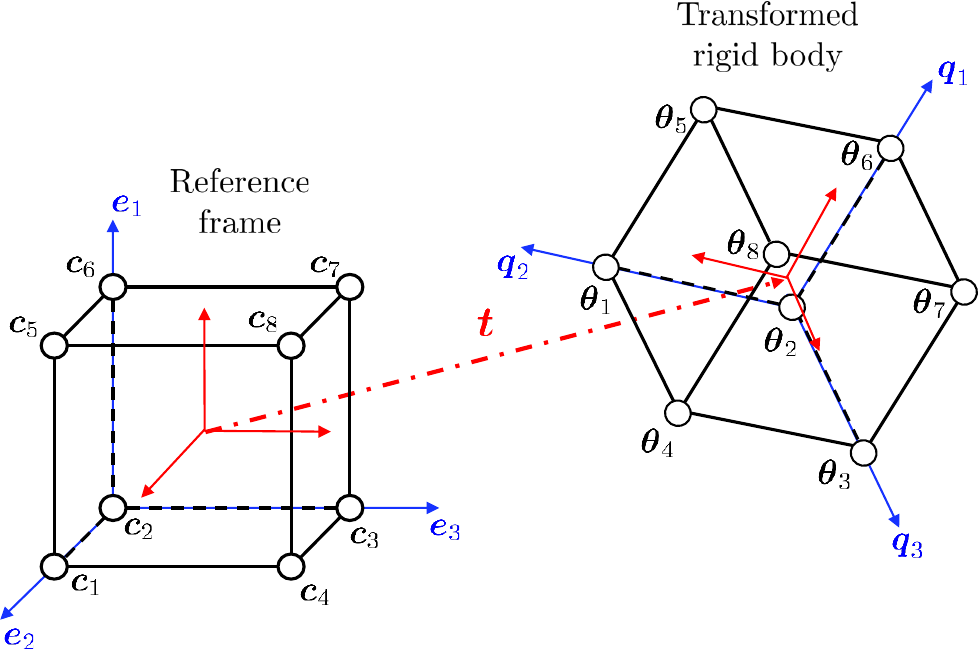}    
\vspace{-4ex}
\caption{Illustration of a rigid body at {\color{black} an initial location aligned with a reference frame $\bm{C}$}, and at another location $\bm{\Theta}$, {\color{black}as} obtained after the linear transformation {\color{black}described in} equation \eqref{eq:basic_model_one_body}, {\color{black}comprising a} \ac{3D} rotation {\color{black}via the matrix} $\bm{Q}$ and {\color{black}a} translation {\color{black} by the} vector $\bm{t}$. Notice that by force of the latter, the rotation matrix $\bm{Q}$ reduces to an identity $\bm{I}=[\bm{e}_1, \bm{e}_2, \bm{e}_3]$ if the body is at its canonic location such that ${\color{black}\bm{\Theta}}=\bm{C}$.}
\label{fig:RB_tra}
\vspace{1ex}
{{\includegraphics[width=\columnwidth]{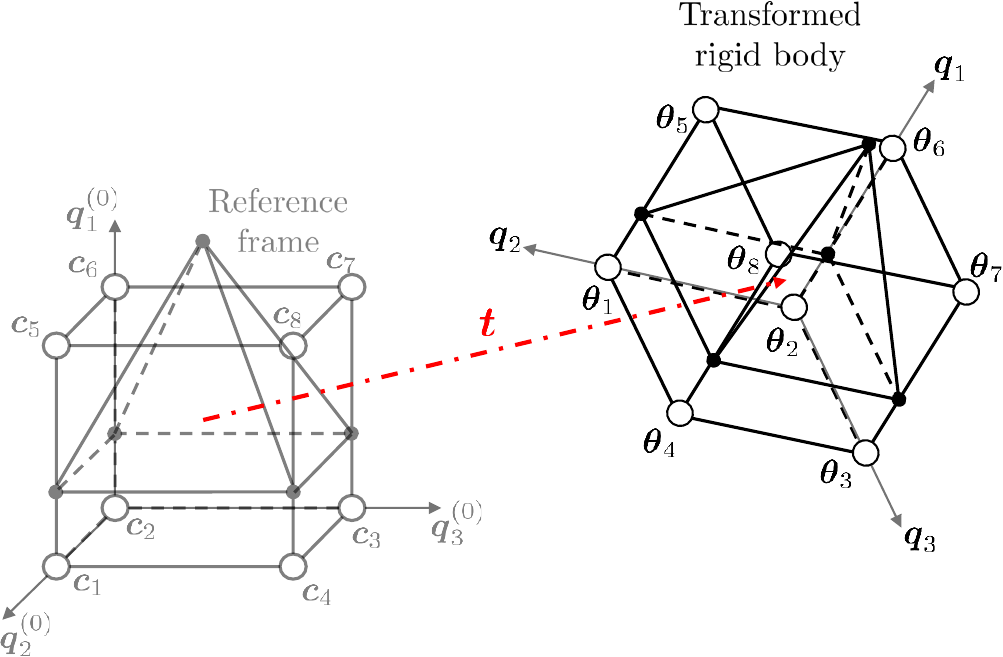}}}
\vspace{-4ex}
\caption{Illustration of a scenario where a rigid body with {\color{black}${\color{black}n_A} = 5$} landmark points seeks to locate another rigid body with {\color{black}${\color{black}n_T} = 8$} points.}
\label{fig:sys}
\vspace{-2ex}
\end{figure}

Before proceeding with the proposed \ac{CRLB} derivation, we first introduce the \ac{RBL} system model and information-centric \ac{FIM} formulation which {\color{black}sets} the basis of subsequent derivations.
To that extend, let us refer to Figure \ref{fig:RB_tra} and consider an arbitrary rigid body in the \ac{3D} space, that can be represented by a collection of $N$ landmark points $\bm{c}_n\in\mathbb{R}^{3\times 1}$, with $n=\{1,\cdots,N\}$, such that the corresponding conformation matrix $\bm{C}$ describing the shape of the body is constructed by the column-wise collection of the individual landmark position vectors $\bm{c}_n$, $i.e.$, $\bm{C}\triangleq[\bm{c}_{1},\cdots,\bm{c}_{N}]\in \mathbb{R}^{3\times N}$.

As also illustrated in Figure \ref{fig:RB_tra}, let ${\color{black}\bm{\Theta}}$ be the location of said {\color{black}transformed} rigid body relative to the {\color{black}conformation matrix} $\bm{C}$, be given by the following linear transformation of the latter
\begin{equation}
\label{eq:basic_model_one_body}
{\color{black}\bm{\Theta}}=\bm{Q}\cdot\bm{C}+\bm{t}\cdot\bm{1}_{N}^{\intercal},
\end{equation}
where $\bm{t}\in\mathbb{R}^{3\times 1}$ is a translation vector given by the distance {\color{black} between the geometric center (centroid) of the body and the centroid in the corresponding reference frame}, $\bm{1}_{N}$ is a column vector with $N$ entries all equal to $1$, and $\bm{Q}\in \mathbb{R}^{3\times 3}$ is a rotation matrix\footnote{The rotation matrix $\bm{Q}$ is part of the special orthogonal group such that $SO(3)=\left\{\bm{Q} \in \mathbb{R}^{3 \times 3}: \bm{Q}^\intercal \bm{Q} = \mathbf{I}, ~\mathrm{det}(\bm{Q})=1\right\}$ \cite{Diebel2006}, which is interpreted as estimating of the $9$ elements of the corresponding rotation matrix $\bm{Q}$ as a whole. However, as shown in \cite{vizitivWCNC2025,Nic_6D_RBL} this representation can be extended by replacing the estimation of $\bm{Q}$ with the estimation of the associated and fundamental yaw, pitch and roll angles $(\alpha, \beta, \gamma)$.}, determined by the corresponding yaw, pitch and roll angles $\alpha, \beta$ and $\gamma$, respectively, namely
\vspace{-1ex}
\begin{eqnarray}
\label{eq:rot_matrix}
\bm{Q} \triangleq \bm{Q}_{z}(\gamma)\,\bm{Q}_{y}(\beta)\,\bm{Q}_{x}(\alpha)&&\\
&&\hspace{-28ex}
=\!\!\!\left[
\begin{array}{@{}c@{\;\,}c@{\;\,}c@{}}
\cos\gamma&-\sin\gamma& 0\\
\sin\gamma& \cos\gamma& 0\\
0 	    & 0           & 1\\
\end{array}\right]\!\!\!\cdot\!\!\!
\left[
\begin{array}{@{}c@{\;\,}c@{\;\,}c@{}}
\cos\beta & 0           & \sin\beta\\
0			& 1			  & 0\\
-\sin\beta& 0 		  & \cos\beta\\
\end{array}\right]\!\!\!\cdot\!\!\!
\left[
\begin{array}{@{\,}c@{\;\,}c@{\;\,}c@{\!}}
1 			& 0			  & 0\\
0			& \cos\alpha& -\sin\alpha\\
0			& \sin\alpha& \cos\alpha\\
\end{array}\right]\nonumber\\
&&\hspace{-28ex}
=\!\!\text{\scalebox{0.75}{$
\left[
\begin{array}{@{\,}c@{\;\,}c@{\;\,}c@{\,}}
\cos\beta\cos\gamma & \sin\alpha\sin\beta\cos\gamma- \cos\alpha\sin\gamma & \cos\alpha\sin\beta\cos\gamma+\sin\alpha\sin\gamma\\
\cos\beta\sin\gamma & \sin\alpha\sin\beta\sin\gamma+ \cos\alpha\cos\gamma & \cos\alpha\sin\beta\sin\gamma-\sin\alpha\cos\gamma\\
-\sin\beta			& \sin\alpha\cos\beta								  & \cos\alpha\cos\beta\\
\end{array}\right]$}}\nonumber\\
&&\hspace{-28ex}
=\!\!\left[
\begin{array}{@{\,}c@{\;\,}c@{\;\,}c@{\,}}
q_{1,1} & q_{1,2} & q_{1,3}\\
q_{2,1} & q_{2,2} & q_{2,3}\\
q_{3,1} & q_{3,2} & q_{3,3}\\
\end{array}\right]\!\!.\nonumber
\end{eqnarray}

{\color{black}As should be clear from equation \eqref{eq:basic_model_one_body}, the translation vector that describes the motion of rigid body from $\bm{C}$ to $\bm{\Theta}$ is given by the vector defined by the centroids of the body at those locations.
But since the centroid of any collection of points is uniquely defined, the same principle can be used to define the translation vector $\bm{t}$ more generally.
Consider therefore the scenario illustrated in Figure \ref{fig:sys}, where} a target rigid body is to be localized by a set of anchor nodes\footnote{Without loss of generality, we will refer to the target body as the first body, and the set of anchors as the second body, where the second body containing the anchor nodes can either be assumed to be multiple nodes of a positioning infrastructure \cite{Chen_2015}, or an actual second rigid body that egoistically seeks to locate the first \cite{Fuehrling_2025}{\color{black}, without any knowledge about the target's conformation}.}, whose locations are assumed to be known.
For the sake of exposition, the set of anchors can be treated as a second rigid body, such that one can speak of two distinct rigid bodies hereafter referred to by their indices $i=\{1,2\}$, which generally have different shapes and/or are characterized by generally distinct numbers ${\color{black}n_T}$ and ${\color{black}n_A}$ of landmark points, respectively.
It follows that, under a common absolute reference frame, the two bodies can be represented by the corresponding distinct conformation matrices $\bm{C}_1\in\mathbb{R}^{3\times {\color{black}n_T}}$ and $\bm{C}_2\in\mathbb{R}^{3\times {\color{black}n_A}}$, with $\bm{C}_i=[\bm{c}_{i,1},\cdots,\bm{c}_{i,N_i}]\in \mathbb{R}^{3\times N_i}$, where $\bm{c}_{i,n}$, is the location of the $n$-th point of the $i$-th body.
\vspace{-3ex}

It is worth emphasizing that {\color{black}the location and orientation of one body relative to another can still be modeled by equation \eqref{eq:basic_model_one_body}, although in general scenarios the two rigid bodies are not identical, such that $\bm{C}_1 \neq \bm{C}_2$}, in the sense that the following still holds
\begin{equation}
\label{eq:basic_model_two_bodies}
{\color{black}\bm{\Theta}}_i={\color{black}\bm{Q}}\cdot\bm{C}_i+{\color{black}\bm{t}}\cdot\bm{1}_{N_i}^{\intercal}, \forall\,i\in\{1,2\},
\end{equation}
where the same rotation matrix $\bm{Q}$ and translation vector $\bm{t}$ is used for both $i\in\{1,2\}$, as illustrated in Figure \ref{fig:sys}.

\subsection{Fisher Information Matrix Construction Framework}
\label{sec:FIM_Framework}

In light of the above, we hereafter consider a general scenario in which two distinct rigid bodies, hereafter referred to by their indices $i=\{1,2\}$, have different shapes and/or are characterized by distinct numbers ${\color{black}n_T}$ and ${\color{black}n_A}$ of landmark points, such that there are $N={\color{black}n_T}+{\color{black}n_A}$ landmark points in total.
Consequently, the {\color{black}indices of the} rigid bodies {\color{black}landmark points} can be referred to as a target group $\mathcal{T}=\{1,\cdots,t,\dots,n_T\}$ and an anchor group $\mathcal{A}=\{1,\cdots,a,\cdots,n_A\}$, respectively, such that the coordinates of the target and anchor groups are represented by the corresponding rigid body conformation matrices, which can be written as $\bm{\Theta}_{\mathcal{T}}\in \mathbb{R}^{3\times n_T}$ and $\bm{\Theta}_{\mathcal{A}}\in \mathbb{R}^{3\times n_A}$, respectively.

The fundamental limits of \ac{RBL} depend on the both type of measurements available and the corresponding errors affecting them, where we emphasize that the type of measurements (distances, angles, etc.) relates to the geometry of the estimation problem, while the measurement errors relates to the statistical uncertainty over the latter quantities.
It will prove convenient, therefore, to separate these two distinct aspects of the input.
To that end, let us start by defining a mapping function that captures the dissimilarity between any arbitrary target point $n$ and reference anchor node $a$, which is given by
\begin{equation}
g(\bm{\theta}_n|\bm{\theta}_a):\mathbb{R}^\eta \rightarrow \mathbb{R},
\end{equation}
where $\eta$ is the dimension of the target and anchor nodes coordinates {\color{black}$\bm{\theta}_n$ and $\bm{\theta}_a$}. 

We shall refer to the mapping function $g(\bm{\theta}_n|\bm{\theta}_a)$ as the \textit{dissimilarity function}, since it yields a measure of dissimilarity between a target {\color{black}node} and anchor node.
The acquired values of the dissimilarity measurements are, however, typically subject to errors, such that we also introduce the corresponding probability density function, hereafter referred to as the \textit{variation function}, which is denoted by {\color{black}$p_{na}(r_{na};g(\bm{\theta}_n|\bm{\theta}_a))$}, where $r_{na}$ is the random variable associated with the measured dissimilarity between the target point $t$ and anchor node $n$.

\begin{table*}
\centering
\caption[Dissimilarity functions and information gradients]{Dissimilarity Functions and Information Gradients in {\color{black}Point-Target} Wireless Localization Systems.}
\label{TAB_DissimilarityFunctions}
\begin{adjustbox}{width=\textwidth}
\begin{threeparttable}
\begin{tabular}{c|c|c}
&&\\[-4ex]
\hline 
&&\\
\bfseries{Measure Type} & \bfseries{Dissimilarity Function $g(\bm{\theta}_n\vert \bm{\theta}_a)$} & \bfseries{Information Gradient} $\mathbf{v}_{na} = \nabla_{_{\!\!\bm{\theta}}} g(\bm{\theta}_n\vert \bm{\theta}_a)$
\\[1.5ex]
\hline
\hline 
&&\\
Distance & $g(\bm{\theta}_n\vert \bm{\theta}_a)=\|\bm{\theta}_n-\bm{\theta}_a\| = d_{na}$ & $\frac{\partial g(\bm{\theta}_n\vert \bm{\theta}_a)}{\partial\bm{\theta}_n}  = \frac{1}{\|\mathbf{d}_{na}\|}[(x_t - x_n),(y_t-y_n),(z_t-z_n)]^{\textup{T}} = \dfrac{\mathbf{d}_{na}}{{d}_{na}}$\\[2ex]
\hline 
&&\\
Angle of Arrival &
$g(\bm{\theta}_n\vert \bm{\theta}_a;\{\mathbf{a},\mathbf{b}\}) =
\acos\!\left(\frac{\langle\mathbf{d}_{na},\mathbf{b}\rangle}{\left\|\mathbf{d}_{na}-\langle\mathbf{d}_{na},\mathbf{a}\rangle\mathbf{a}\right\|}\right)$ & $\frac{\partial g(\bm{\theta}_n\vert \bm{\theta}_a)}{\partial\bm{\theta}_n} =\frac{1}{\|\mathbf{d}_{na}\|^2}[(y_t - y_n),-(x_t-x_n),0]^{\textup{T}}={\tiny\left[\begin{array}{@{}r@{\;\;}r@{\;\;}r@{\,}}
0 & 1 &0\\
-1 & 0 &0\\
0 & 0 &0
\end{array}\right]}\dfrac{\mathbf{d}_{na}}{{d}^2_{na}}$
\\[2ex]
\hline 
&&\\ 
Angle Difference of Arrival & 
$g(\bm{\theta}_n\vert \bm{\theta}_a,\bm{\theta}_k)= \acos\!\left(\frac{d_{na}^2+d_{ka}^2-d_{nk}^2}{2d_{na}d_{ka}}\right)$ & $\frac{\partial g(\bm{\theta}_n\vert \bm{\theta}_a,\bm{\theta}_k)}{\partial\bm{\theta}_n} =\frac{d_{na}}{\|\mathbf{d}_{nk} \times  \mathbf{d}_{na}\|} (\rho_1 \mathbf{d}_{na} +\rho_2 \mathbf{d}_{nk})$
\\[3ex]
\hline 
\end{tabular}
\begin{tablenotes}
\item[*]Notes:
a) The coordinate vectors are explicitly written as $\bm{\theta}_a \triangleq [x_n, y_n, z_n]^\text{T}$ and $\bm{\theta}_n \triangleq [x_t, y_t, z_t]^\text{T}$; b) The distance vector between $\bm{\theta}_n$ and $\bm{\theta}_a$ is defined as $\mathbf{d}_{na} \triangleq \bm{\theta}_n - \bm{\theta}_a$; c) The notation $\langle \cdot,\cdot\rangle$ denotes the inner product between two vectors; d) The unit-norm norm vector $\mathbf{a}$, centered at $\bm{\theta}_a$, defines a reference plane for angle-of-arrival measurements; and d) The unit-norm vector $\mathbf{b}$, also centered at $\bm{\theta}_a$ and belonging to the latter plane, is a reference vector against which the angle of arrival is measured (see Figure \ref{Fig:IllustrationAngleDissimilarity}); e) The coefficients $\rho_1 \triangleq \frac{d_{ka}^2-(d_{na}^2+d_{nk}^2)}{2d_{ka}d_{na}}$ and $\rho_2 \triangleq \frac{1}{d_{ka}}$.
\end{tablenotes}
\end{threeparttable}
\end{adjustbox}
%
\vspace{1ex}
\caption[Information intensities for different variation functions]{Information Intensities for Different Variation Functions}
\label{TAB_InformationIntensities}
\begin{adjustbox}{width=\textwidth}
\begin{threeparttable}
\begin{tabular}{c|c|c|c}
&&&\\[-4ex]
\hline 
&&&\\
\bfseries{Model} & \bfseries{Variation Function $p(r_{na};g(\bm{\theta}_n\vert \bm{\theta}_a))$} & \bfseries{Information Intensity $\sqrt{F}$} & \bfseries{Reference}
\\
\hline
\hline 
&&&\\
Normal  & $p(r_{na};g(\bm{\theta}_n\vert \bm{\theta}_a)) = \mathcal{N}(g(\bm{\theta}_n\vert \bm{\theta}_a),\sigma^{2})$ & $\sqrt{F}=\frac{1}{\sigma}$ & \cite{PatwariTSP2003}\\[1ex]
\hline 
&&&\\
Normal with $\sigma^2 = \beta(g(\bm{\theta}_n\vert \bm{\theta}_a))^{\alpha}$$^{^1}$ & $p(r_{na};g(\bm{\theta}_n\vert \bm{\theta}_a)) =\mathcal{N}(g(\bm{\theta}_n\vert \bm{\theta}_a),\beta(g(\bm{\theta}_n\vert \bm{\theta}_a))^{\alpha})$ & $\sqrt{F}=\sqrt{\frac{1}{\beta (g(\bm{\theta}_n\vert \bm{\theta}_a))^\alpha} + \frac{\alpha^{2}}{2(g(\bm{\theta}_n\vert \bm{\theta}_a))^{2}}}$ & \cite{BuehrerMILCOM2008}\\[2ex]
\hline 
&&&\\
Von-Mises & $p(r_{na};g(\bm{\theta}_n\vert \bm{\theta}_a)) = \dfrac{\mathrm{exp}\left(\omega\,\mathrm{cos}(r_{na}-g(\bm{\theta}_n\vert \bm{\theta}_a))\right)}{2\pi I_{0}(\omega)}$ & $\sqrt{F}=\frac{\omega}{\sqrt{2}}$ & Appendix \ref{SEC_VonMisesFisher}\\[2.5ex]
\hline 
&&&\\
Nakagami-$m$  with controlling spread {$\Upsilon^{^{2}}$}&  $p(r_{na};g(\bm{\theta}_n\vert \bm{\theta}_a)) = \dfrac{2m^{m}}{\Gamma(m) \Upsilon^{m}} r_{na}^{2m-1} \exp{\left(-\dfrac{m}{\Upsilon}r_{na}^2\right)}$ & {\color{black}$\sqrt{F}=\sqrt{\frac{m(4m - 3)}{\Upsilon(m - 1)}}$} & Appendix \ref{SEC_NakagamiFisher}\\[2.5ex]
\hline 
&&&\\
Gamma  with shape and scale parameter $\kappa$ and $\upsilon^{^{3}}$&  $p(r_{na};g(\bm{\theta}_n\vert \bm{\theta}_a)) = \dfrac{1}{\Gamma(\kappa) \upsilon^{\kappa}} r_{na}^{\kappa-1} \exp{\left(-\dfrac{r_{na}}{\upsilon}^2\right)}$ & $\sqrt{F}=\sqrt{\frac{1}{\upsilon^2(\kappa-2)}}$ & Appendix \ref{SEC_GammaFisher}\tabularnewline[2.5ex]
\hline 
\end{tabular}
\begin{tablenotes}
\item[1]  Here $\alpha$ is the pathloss exponent, while $\beta$ is a scaling factor related to the sensitivity of the transmitter.
\item[2] Here $m \geq \tfrac{1}{2}$ and $\Upsilon>0$ are the shape and spread control  parameters. 
\item[3] Here $\kappa>0$ and $\upsilon>0$ are the shape and scale parameter.
\end{tablenotes}
\end{threeparttable}
\end{adjustbox}
\vspace{-3ex}
\end{table*}

Some examples of {\color{black}point-target} dissimilarity functions are given in Table \ref{TAB_DissimilarityFunctions}, with a few corresponding suitable variation functions {\color{black}for different error models} given in Table \ref{TAB_InformationIntensities}.

In possession of a given dissimilarity and variation function associated with the Fisher information between target $t$ and anchor $n$ given the variates $r_{na}$, is described by \cite{Rao_92}
\begin{equation}
\label{eq:Fisher_Info}
F_{na}=\mathbb{E}_{r_{na}}\left[\left(\frac{\partial \ln p_{na}(r_{na};g(\bm{\theta}_n\vert \bm{\theta}_a))}{\partial g(\bm{\theta}_n\vert \bm{\theta}_a)}\right)^{2}\right],
\end{equation}
where $\mathbb{E}[\cdot]$ denotes the expectation operator, {\color{black}applied to} the gradient of the log-likelihood function with respect to the dissimilarity function $g(\bm{\theta}_n|\bm{\theta}_a)$.

Taking into account all pairs of target point and anchor node, collected in the set $\mathcal{P}\triangleq\{(n,a),\forall n \text{ and } \forall a\}$, the likelihood function corresponding to all target points is the group $\bm{\Theta}_{\mathcal{T}}$ can then be given by
\begin{equation}
\label{eq:LogLikelihood}
\mathcal{L}(\mathbf{r}|\bm{\Theta}_{\mathcal{T}})=\prod_{(n, a) \in \mathcal{P}}p_{na}(r_{na};g(\bm{\theta}_n|\bm{\theta}_a)),
\vspace{-1ex}
\end{equation}
where $\mathbf{r}$ is the vector of all {\color{black}independent} measurements, and $\bm{\Theta}_{\mathcal{T}}$ is the vector of all target coordinates.

Collecting the expression above for all target points, the \ac{FIM} can be constructed as
\begin{equation}
\label{eq:conv_FIM}
\mathbf{F}_{\bm{\Theta}_{\mathcal{T}}}=\mathbb{E}[\nabla_{\bm{\Theta}_{\mathcal{T}}}\ln(\mathcal{L}(\mathbf{r}|\bm{\Theta}_{\mathcal{T}}))\nabla^{\intercal}_{\bm{\Theta}_{\mathcal{T}}}\ln(\mathcal{L}(\mathbf{r}|\bm{\Theta}_{\mathcal{T}}))],
\end{equation}
where $\nabla_{\bm{\Theta}_{\mathcal{T}}}$ is the gradient operator with respect to the target coordinates $\bm{\Theta}_{\mathcal{T}}$.

We refer to the \ac{FIM} construction directly via equation \eqref{eq:conv_FIM} as \emph{element-centric}, since in this case the matrix $\mathbf{F}_{\bm{\Theta}_{\mathcal{T}}}$ is obtained on an element-by-element basis from equation \eqref{eq:Fisher_Info}, and remark that although this usual approach to construct the \ac{FIM} is concise mathematically, it is not very insightful in elucidating the contribution of each measurement towards knowledge on the parameters to be estimated.

In contrast, we revise in the sequel an alternative approach to construct \acp{FIM} \cite{Ghods_2014} in which $\mathbf{F}_{\bm{\Theta}_{\mathcal{T}}}$ is obtained by a sum of rank-one matrices, each carrying the contribution of only one measurement between an anchor node $a$ and a target point $n$, of whatever type.

\begin{lemma}[Sum-product Formulation of \ac{FIM}]
\label{lm:InfocentricFIM}
Consider the unbiased estimation of the parameters of interest in $\bm{\Theta}_{\mathcal{T}}$ from the input vector $\mathbf{r}$ \textit{with the associated dissimilarity function} $g(\bm{\theta}_n | \bm{\theta}_a)$ \textit{and variation function} $p_{na}(r_{na}; g(\bm{\theta}_n | \bm{\theta}_a))$, \textit{where} $n \in \{1, \dots, n_T\}$ \textit{and} $a \in \{1, \dots, N\}$. Then the \ac{FIM} $\mathbf{F}_{\bm{\Theta}_{\mathcal{T}}}$ can be obtained in an information-centric approach from the sum of information vector products \cite{Velde_14}:
\begin{equation}
\label{eq:new_FIM}
\mathbf{F}_{\bm{\Theta}_{\mathcal{T}}} = \sum_{(n, a) \in \mathcal{P}} \mathbf{u}_{na} \mathbf{u}_{na}^{T} = \sum_{(n, a) \in \mathcal{P}} \lambda_{na} \mathbf{v}_{na} \mathbf{v}_{na}^{T},
\vspace{-1ex}
\end{equation}
where $\mathbf{u}_{na}$ is referred to as the information vector, given by
\begin{equation}
\label{eq:info_vector}
\mathbf{u}_{na}\triangleq\frac{\partial g(\bm{\theta}_n|\bm{\theta}_a)}{\partial \mathrm{vec}(\bm{\Theta}_{\mathcal{T}})}\sqrt{F_{na}}=\sqrt{\lambda_{na}}\mathbf{v}_{na},
\end{equation}
while $\sqrt{F_{na}}=\sqrt{\lambda_{na}}$ is the information intensity that captures impact of the errors that the measurement $r_{na}$ is subject to, and $\mathbf{v}_{na}\triangleq\frac{\partial g(\bm{\theta}_n|\bm{\theta}_a)}{\partial \mathrm{vec}(\bm{\Theta}_{\mathcal{T}})}$ being the information gradient that captures the impact of the type of measurement $r_{na}$, onto the \ac{FIM}.
\end{lemma}

\textbf{Proof:} Let us start by noticing that given equations \eqref{eq:Fisher_Info}, \eqref{eq:LogLikelihood} and \eqref{eq:info_vector}, equation \eqref{eq:conv_FIM} can be rewritten as 
\begin{subequations}
\label{eq:FIM_with_score}
\begin{eqnarray}
\mathbf{F}_{\bm{\Theta}_{\mathcal{T}}} = \hspace{-4ex}&&\mathbb{E} \bigg[ \sum_{na} \sum_{ij} \mathbf{U}_{na} (r_{na} | \bm{\Theta}_{\mathcal{T}}) \mathbf{U}^{T}_{ij} (r_{ij} | \bm{\Theta}_{\mathcal{T}}) \bigg]\\
=\hspace{-4ex}&& \sum_{na} \mathbb{E} \left[ \mathbf{U}_{na} (r_{na} | \bm{\Theta}_{\mathcal{T}}) \mathbf{U}_{na}^{ \intercal} (r_{na} | \bm{\Theta}_{\mathcal{T}}) \right] \\
&&+ \sum_{\substack{na, ij \\ na \neq ij}} \underbrace{\mathbb{E} [\mathbf{U}_{na} (r_{na} | \bm{\Theta}_{\mathcal{T}})]}_{=0} \underbrace{\mathbb{E} \left[ \mathbf{U}_{ij}^{ \intercal} (r_{ij} | \bm{\Theta}_{\mathcal{T}}) \right]}_{=0},\nonumber\\
= \hspace{-4ex}&&\sum_{na} \mathbb{E} \left[ \mathbf{U}_{na} (r_{na} | \bm{\Theta}_{\mathcal{T}}) \mathbf{U}_{na}^{ \intercal} (r_{na} | \bm{\Theta}_{\mathcal{T}}) \right],
\label{eq:FIM_simpl}
\end{eqnarray}
where $\mathbf{U}_{na} (r_{na} | \bm{\Theta}_{\mathcal{T}})$ is well-known as the \textit{score function} \cite{Cox_1979} of the random variables $r_{na}$, which can be extracted from equation \eqref{eq:conv_FIM} and is defined as
\end{subequations}
\begin{equation}
\mathbf{U}_{na} (r_{na} | \bm{\Theta}_{\mathcal{T}}) \triangleq \nabla_{\bm{\Theta}_{\mathcal{T}}} \ln \left( p_{na} (r_{na}; g(\bm{\theta}_n | \bm{\theta}_a)) \right),
\end{equation}
and we have utilized the fact that for an unbiased estimator the score function has an expected value of zero \cite{Cox_1979}.

Using the chain rule, the score function can be rewritten as
\begin{equation}
\mathbf{U}_{na} (r_{na} | \bm{\Theta}_{\mathcal{T}}) = \frac{\partial \ln p_{na} (r_{na} ; g (\bm{\theta}_n | \bm{\theta}_a))}{\partial g (\bm{\theta}_n | \bm{\theta}_a)}
\frac{\partial g (\bm{\theta}_n | \bm{\theta}_a)}{\partial \bm{\Theta}_{\mathcal{T}}},
\label{eq:score_chain}
\end{equation}
where the first partial derivative term of equation (\ref{eq:score_chain}) is a scalar, while the second is a vector. 

\begin{subequations}
\label{eq:FIM_infocentric}
Combining equations \eqref{eq:FIM_simpl} and \eqref{eq:score_chain} yields
\vspace{-1ex}
\begin{eqnarray}
\mathbf{F}_{\bm{\Theta}_{\mathcal{T}}} = \hspace{-4ex}&&\sum_{n=1}^{n_T} \sum_{a=1}^{n_A} \mathbb{E} \left[ \Big( \frac{\partial \ln p_{na} (r_{na} ; g (\bm{\theta}_n | \bm{\theta}_a))}{\partial g (\bm{\theta}_n | \bm{\theta}_a)} \Big)^{\!2} \right]\times\nonumber \\
&&
\frac{\partial g (\bm{\theta}_n | \bm{\theta}_a)}{\partial \bm{\Theta}_{\mathcal{T}}}\frac{\partial g (\bm{\theta}_n | \bm{\theta}_a)^{ \intercal}}{\partial \bm{\Theta}_{\mathcal{T}}}\\
=\hspace{-4ex}&& \sum_{n=1}^{n_T} \sum_{a=1}^{n_A} F_{na} \frac{\partial g (\bm{\theta}_n | \bm{\theta}_a)}{\partial \bm{\Theta}_{\mathcal{T}}} \frac{\partial g (\bm{\theta}_n | \bm{\theta}_a)^{ \intercal}}{\partial \bm{\Theta}_{\mathcal{T}}},\\
=\hspace{-4ex}&&\sum_{n=1}^{n_T} \sum_{a=1}^{n_A} \mathbf{u}_{na} \mathbf{u}_{na}^{ \intercal} = \sum_{n=1}^{n_T} \sum_{a=1}^{n_A} \lambda_{na} \mathbf{v}_{na} \mathbf{v}_{na}^{ \intercal}\in\mathbb{R}^{\eta n_T \times \eta n_T},\nonumber\\[-2ex]
&&\label{eq:FIM_Theta}\\[-4ex]
&&\nonumber
\end{eqnarray}
where the quantity $F_{na}$ is the Fisher information of the input $r_{na}$ given in equation \eqref{eq:Fisher_Info}, while $\mathbf{u}_{na}$ and $\mathbf{v}_{na}$ are the information vector and the information gradient defined in the statement of Lemma \ref{lm:InfocentricFIM}.
\end{subequations} \hfill$\qed$

{\color{black}Although evident from the proof itself, we emphasize that the equivalence between equations \eqref{eq:conv_FIM} and \eqref{eq:new_FIM} is conditioned on the assumption of independence among the measurements in the input vector, and remark that such an assumption does not detract from the result, since it implies a maximization of the Fisher information \cite{StevenKay1993}, leading to a corresponding \ac{CRLB} with the lowest possible values ($i.e.$, true fundamental limit).
Finally, we highlight that in turn, the alternative \ac{FIM} expression in equation \eqref{eq:new_FIM} offers, beyond elegance, new insight by enabling easily access and assessment of the contribution of any given measurement $r_{na}$.}

Examples for the information gradient $\mathbf{v}_{na}$ and the information intensity $\lambda_{na}$ for different types of measurements are given in Table \ref{TAB_DissimilarityFunctions} and Table \ref{TAB_InformationIntensities},
and we emphasize that although the results of constructing the \ac{FIM} using equations \eqref{eq:conv_FIM} and \eqref{eq:FIM_infocentric} are identical, the latter is a lot more convenient for extracting insight about the impact of each measurement on the \ac{FIM}.

In particular, as pointed out already in \cite{Ghods_2014}, it is easy to study the consequence of removing the contribution of any given measurement $r_{na}$ from the \ac{FIM} constructed via equation \eqref{eq:FIM_infocentric}, or investigate the impact of higher or smaller errors on any such variates.

In what follows, however, we take further advantage of the convenient structure of equation \eqref{eq:FIM_infocentric} to derive a \ac{FIM} and latter the corresponding \ac{CRLB} for \ac{RBL} estimators, whereby the fundamental limit on the estimates not of individual landmark points, but rather of the overall location and orientation of the rigid body is of interest.


\vspace{-2ex}
\section{Derivation of the CRLB for RBL}
\label{sec:prop}

As discussed in Subsection \ref{sec:RBL}, the objective of \ac{RBL} algorithms \cite{Chen_2015,vizitivWCNC2025,Nic_6D_RBL} is to find the position, as determined by the translation vector $\bm{t}$, and the orientation, as determined by the yaw, pitch and roll angles $\alpha, \beta$ and $\gamma$, respectively, captured by the corresponding rotation matrix $\bm{Q}$, as described by equation \eqref{eq:basic_model_two_bodies} and illustrated in Figure \ref{fig:sys}.
In order to assess the performance of \ac{RBL} algorithms it is therefore relevant to obtain \acp{FIM} and corresponding \acp{CRLB} relative to the estimation parameters $\bm{t}$ and $\bm{Q}$.

\vspace{-2ex}
\subsection{CRLB Formulation for the Translation Vector}

Due to the independence among the elements of the translation vector, and among the yaw, pitch and roll angles, captured in the rotation matrix, the bounds for both parameters can be obtained individually.
To that end, let us first consider the previously described \ac{RBL} scenario and map the notation to what has been introduced in the construction of the \ac{FIM}.

In particular, in a \ac{RBL} scenario the target group $\bm{\Theta}_{\mathcal{T}}$ can be defined as the positions of the transformed rigid body, {\color{black}as described before}, such that equation \eqref{eq:basic_model_two_bodies} can be {\color{black}specifically} rewritten as
\begin{equation}
\label{eq:basic_model_one_body_i}
\bm{\Theta}_{\mathcal{T}}=\bm{Q}\cdot\bm{C}_{1}+\bm{t}\cdot\bm{1}_{n_T}^{\intercal},
\vspace{-0.5ex}
\end{equation}
which, for convenience, can be rewritten for the individual target point $n$ as
\begin{equation}
\label{eq:basic_model_one_body_i_one_sensor}
\bm{\theta}_{n}=\bm{Q}\cdot\bm{c}_n+\bm{t}.
\vspace{-0.5ex}
\end{equation}

Following the information-centric construction method described earlier, the \ac{FIM} for the translation vector $\bm{t}$ (in three dimensions) is then given by
\begin{equation}
    \color{black}
\label{eq:FIM_t1}
\begin{split}
\mathbf{F}_{\bm{t}}&=\sum_{(n, a) \in \mathcal{P}}\lambda_{na}\mathbf{v}_{na}\mathbf{v}_{na}^{\intercal}\\
&=\sum_{(n, a) \in \mathcal{P}}\lambda_{na}\frac{\partial g(\bm{\theta}_n|\bm{\theta}_a)}{\partial \bm{t}}\Big(\frac{\partial g(\bm{\theta}_n|\bm{\theta}_a)}{\partial \bm{t}}\Big)^\intercal\in \mathbb{R}^{3\times3}.   
\end{split}
\end{equation}

The information gradient associated with the dissimilarity function $g(\bm{\theta}_n|\bm{\theta}_a)$ w.r.t the translation vector $\bm{t}$ in the latter equation can be obtained by using double dot product\footnote{Note that even though it seems intuitive to use the chain rule to decompose the gradient into two parts, namely a partial derivative of the dissimilarity function w.r.t the matrix of target coordinates $\bm{\Theta}_\mathcal{T}$, and a partial derivative of $\bm{\Theta}_\mathcal{T}$ w.r.t the translation vector $\bm{t}$, this concept cannot be applied directly, due to the {\color{black}mismatching sizes of the partial derivatives}.}, also known as the Frobenius inner product and the differential approach\cite{Magnus_2019,Herzog_2025}.
As described in Appendix \ref{sec:gradient}, depending on the type of measurement, the information gradient changes and thus, the \ac{FIM} for the translation vector $\bm{t}$ with respect to distance, \ac{AoA} and \ac{ADoA} measurements, collected in the sets $\mathcal{P}_d$, $\mathcal{P}_\psi$ and $\mathcal{P}_{\psi'}$, respectively, can be expressed as
\begin{eqnarray}
\label{eq:FIM_t_final}
\mathbf{F}_{\bm{t}}=\hspace{-4ex}&&\!\!\!\!\sum_{(n, a) \in \mathcal{P}_d}\!\!\!\!\!\lambda_{na}g' \big|_{\mathbf{t}}^{d}\Big(g' \big|_{\mathbf{t}}^{d}\Big)^\intercal+\!\!\!\!\!\sum_{(n, a) \in \mathcal{P}_\psi}\!\!\!\!\!\lambda_{na}g' \big|_{\mathbf{t}}^{\psi}\Big(g' \big|_{\mathbf{t}}^{\psi}\Big)^\intercal\nonumber\\
&&+\!\!\!\!\!\!\sum_{(n, a) \in \mathcal{P}_{\psi'}}\!\!\!\!\!\lambda_{na}g' \big|_{\mathbf{t}}^{\psi'}\Big(g' \big|_{\mathbf{t}}^{\psi'}\Big)^\intercal\in \mathbb{R}^{\eta\times\eta},
\end{eqnarray}
where $\lambda_{na}$ is the information intensity given in Table \ref{TAB_InformationIntensities} and the variables $g' \big|_{\mathbf{t}}^{d}$, $g' \big|_{\mathbf{t}}^{\psi}$ and $g' \big|_{\mathbf{t}}^{\psi'}$ denote the gradients of the dissimilarity function w.r.t the translation vector $\bm{t}$, {\color{black}for} the distance, angle of arrival and angle difference of arrival {\color{black}measurements}, respectively.

We emphasize the simplicity of the \ac{CRLB} for the translation vector given by this expression compared to what one would obtain working directly with equation \eqref{eq:conv_FIM}, which is a consequence of the information-centric approach to construct \acp{FIM}, as described in Subsection \ref{sec:FIM_Framework}, and the fact that translation vector $\bm{t}$ is a coordinate vector similar to $\bm{\theta}_n$.

With the \ac{FIM} in hand, the \ac{CRLB} for the translation vector is a lower bound on the variance of any unbiased estimator, such that given a corresponding estimate $\hat{\bm{t}}$, we have\cite{kay1998detection}
\begin{equation}
\label{eq:CRLB_Matrix}
\bm{\Omega}_{\bm{t}}\triangleq\mathbb{E}\Big[(\hat{\bm{t}}-\bm{t})(\hat{\bm{t}}-\bm{t})^\intercal\Big]\succeq\bm{F}^{-1}_{\bm{t}}.
\end{equation}

In many cases, one is not interested in the positive-semidefine inequality form of the \ac{CRLB} given in equation \eqref{eq:CRLB_Matrix}, but on the \ac{CRLB} for $\bm{t}$ as whole, which is given by
\begin{equation}
    \color{black}
    \label{eq:t_CRLB}
\bar{\omega}_{\bm{t}}\triangleq \frac{\text{tr}(\bm{F}^{-1}_{\bm{t}})}{\eta}.
\end{equation}


\vspace{-2ex}
\subsection{CRLB Formulation for the Rotation Matrix}

While in some applications the yaw, pitch and roll angles $\alpha$, $\beta$ and $\gamma$ are of interest, in the case of \ac{RBL} the rotation matrix $\bm{Q}$ is usually of interest, since instead of estimating the angles individually, the rotation matrix $\bm{Q}$ can be estimated directly.

Following the same approach as before, the \ac{FIM} for the rotation matrix $\bm{Q}$ can be derived by using the differential approach the different dissimilarity functions $g(\bm{\theta}_n|\bm{\theta}_a)$ w.r.t the vectorized version rotation matrix $\text{vec}(\bm{Q})$, which leads to
{ \color{black}
\begin{eqnarray} 
\label{eq:FinalFIMQ}  
\mathbf{F}_{\bm{Q}}\hspace{-4ex}&&=\!\!\!\sum_{(n, a) \in \mathcal{P}}\!\!\!\lambda_{na}\mathbf{v}_{na}\mathbf{v}_{na}^{\intercal}=\!\!\!\sum_{(n, a) \in \mathcal{P}}\!\!\!\lambda_{na}\frac{\partial g(\bm{\theta}_n|\bm{\theta}_a)}{\partial \text{vec}(\bm{Q})}\Big(\frac{\partial g(\bm{\theta}_n|\bm{\theta}_a)}{\partial \text{vec}(\bm{Q})}\Big)^\intercal\nonumber\\
&&=\sum_{(n, a) \in \mathcal{P}_d}\!\!\!\lambda_{na}g' \big|_{\mathbf{Q}}^{d}\Big(g' \big|_{\mathbf{Q}}^{d}\Big)^\intercal+\sum_{(n, a) \in \mathcal{P}_\psi}\!\!\!\lambda_{na}g' \big|_{\mathbf{Q}}^{\psi}\Big(g' \big|_{\mathbf{Q}}^{\psi}\Big)^\intercal\nonumber\\
&&+\sum_{(n, a) \in \mathcal{P}_{\psi'}}\!\!\!\lambda_{na}g' \big|_{\mathbf{Q}}^{\psi'}\Big(g' \big|_{\mathbf{Q}}^{\psi'}\Big)^\intercal\in \mathbb{R}^{\eta^2\times\eta^2},
\end{eqnarray}}
where the information intensity is defined as before, and the information gradients for the different types of measurements are derived in Appendix \ref{sec:gradient}.

Finally, given that the estimator of the rotation matrix $\hat{\bm{Q}}$ is unbiased, the covariance matrix corresponding to the unbiased estimate $\hat{\bm{Q}}$ is bounded by the CRLB \cite{kay1998detection}
\begin{equation}
\bm{\Omega}_{\bm{Q}}\triangleq\mathbb{E}\Big[(\hat{\bm{Q}}-\bm{Q})(\hat{\bm{Q}}-\bm{Q})^\intercal\Big]\succeq\bm{F}^{-1}_{\bm{Q}}.
\end{equation}

The average \ac{CRLB} for the rotation matrix $\bm{Q}$ can be obtained by taking the trace of the inverse of the \ac{FIM} and scaling by the dimension of it, which for the rotation matrix is $\eta^2$, leading to
\vspace{-1.5ex}
\begin{equation}
    \color{black}
{\bar{\omega}_{\bm{Q}}}=\frac{\text{tr}(\bm{F}^{-1}_{\bm{Q}})}{\eta^2}.
\end{equation}

However, while {\color{black}the formulation in equation \eqref{eq:FinalFIMQ}} is valid for an arbitrary matrix of interest, this construction does not take into account the fact that the rotation matrix $\bm{Q}$ is constrained since it belongs to the special orthogonal group, $i.e.$, $\bm{Q}\in\mathcal{SO}(\eta)$, satisfying the orthogonality condition $\bm{Q}^\intercal\bm{Q}=\bm{I}_\eta$ and the determinant condition $\det(\bm{Q})=1$.
To mitigate this issue, the set of constraints {\color{black} in \ac{3D} space} can be defined as 
\begin{equation}
\begin{split}
\mathbf{h}(\bm{Q})= & {\left[\left\|\mathbf{q}_1\right\|^2-1, \mathbf{q}_2^{\top} \mathbf{q}_1, \mathbf{q}_3^{\top} \mathbf{q}_1,\right.} \\
& \left.\left\|\mathbf{q}_2\right\|^2-1, \mathbf{q}_2^{\top} \mathbf{q}_3,\left\|\mathbf{q}_3\right\|^2-1\right]^{\top}=\mathbf{0}_6,
\end{split}
\end{equation}
where $\mathbf{q}_1$, $\mathbf{q}_2$ and $\mathbf{q}_3$ are the columns of the rotation matrix $\bm{Q}$, and $\mathbf{0}_6$ is a zero vector of size $6\times 1$.
Following \cite{Stoica_1998,Nazari_2023}, the constraints can be incorporated into the \ac{FIM} by defining the constraining matrix $\mathbf{M}$, which is given by
\vspace{-1ex}
\begin{equation}
\mathbf{M}=\left[\begin{array}{ccc}
-\mathbf{q}_3 & \mathbf{0}_3 & \mathbf{q}_2 \\
\mathbf{0}_3 & -\mathbf{q}_3 & -\mathbf{q}_1 \\
\mathbf{q}_1 & \mathbf{q}_2 & \mathbf{0}_3
\end{array}\right] \in \mathbb{R}^{9 \times 3},
\end{equation}
such that $\mathbf{G}(\bm{Q}) \mathbf{M}=\mathbf{0}$, where $[\mathbf{G}(\bm{Q})]_{i, j}=\partial[\mathbf{h}(\bm{Q})]_i / \partial[\bm{Q}]_j$.
Finally, the constrained \ac{CRLB} can be found by setting $\mathbf{F}_{\bm{Q}}$ such that 
\vspace{-2ex}
\begin{equation}
    \color{black}
{\bar{\omega}_{\bm{Q}}^{(\mathrm{CCRB})}}=\mathbf{M}\left(\mathbf{M}^{\top} \mathbf{F}_{\bm{Q}} \mathbf{M}\right)^{-1} \mathbf{M}^{\top}.
\end{equation}

{\color{black} The dissimilarity functions and corresponding information gradient vectors with respect to the translation vector t and rotation matrix Q derived in Appendix \ref{sec:gradient}   are summarized in Table \ref{TAB_DissimilarityFunctionsRBL}.

\begin{table*}
\centering
\color{black} 
\caption{Dissimilarity Functions and Information Gradients in RBL Localization Systems.}
\label{TAB_DissimilarityFunctionsRBL}
\begin{adjustbox}{width=\textwidth}
\begin{threeparttable}
\begin{tabular}{c|c|c|c}
&&\\[-4ex]
\hline 
&&\\
\bfseries{Measure Type} & \bfseries{Dissimilarity Function $g(\bm{\theta}_n\vert \bm{\theta}_a)$} & \bfseries{ Information Gradient $\frac{\partial g(\bm{\theta}_n\vert \bm{\theta}_a)}{\partial \bm{t}} $} & \bfseries{ Information Gradient $\frac{\partial g(\bm{\theta}_n\vert \bm{\theta}_a)}{\partial \text{vec}(\bm{Q})} $} 
\\[1.5ex]
\hline
\hline 
&&\\
Squared Distance & $\|\bm{\theta}_n-\bm{\theta}_a\|^2 = d_{na}^2$ & $  2 (\bm{Q}\bm{c}_t + \bm{t} - \bm{\theta}_a)$ & $ 2 (\bm{c}_t \otimes \bm{I}_{\eta}) (\bm{Q}\bm{c}_t + \bm{t} - \bm{\theta}_a)$\\[2ex]
\hline 
&&\\
Angle of Arrival &
$
\acos\!\left(\frac{\langle\mathbf{d}_{na},\mathbf{b}\rangle}{\left\|\mathbf{d}_{na}-\langle\mathbf{d}_{na},\mathbf{a}\rangle\mathbf{a}\right\|}\right)$ & $-\frac{1}{\sqrt{1 - x^2}} \cdot \left(\frac{(\bm{u}^{\intercal} \bm{P} \bm{u} ) \boldsymbol{b} - ( \bm{u}^{\intercal} \boldsymbol{b} )\cdot \bm{P} \bm{u}}{(\|\bm{P} \bm{u}\|)^3}\right)$&$-\frac{(\bm{c}_t \otimes \bm{I}_{\eta})}{\sqrt{1 - x^2}} \cdot \left(\frac{(\bm{u}^{\intercal} \bm{P} \bm{u} ) \boldsymbol{b} - ( \bm{u}^{\intercal} \boldsymbol{b} )\cdot \bm{P} \bm{u}}{(\|\bm{P} \bm{u}\|)^3}\right)$
\\[2ex]
\hline 
&&\\ 
Angle Difference of Arrival & 
$ \acos\!\left(\frac{d_{na}^2+d_{ka}^2-d_{nk}^2}{2d_{na}d_{ka}}\right)$ & $-\frac{1}{\sqrt{1 - x^2}} \cdot \left(\frac{\bm{u}}{2\|\bm{u}\|d} -\frac{\bm{u}d}{2\|\bm{u}\|^3} +\frac{||\bm{v}||^2u}{2\|\bm{u}\|^3d}-\frac{\bm{v}}{\|\bm{u}\|d}\right)$&$-\frac{(\bm{c}_t \otimes \bm{I}_{\eta})}{\sqrt{1 - x^2}} \cdot \left(\frac{\bm{u}}{2\|\bm{u}\|d} -\frac{\bm{u}d}{2\|\bm{u}\|^3} +\frac{||\bm{v}||^2u}{2\|\bm{u}\|^3d}-\frac{\bm{v}}{\|\bm{u}\|d}\right)$
\\[3ex]
\hline 
\end{tabular}
\begin{tablenotes}
\item[*]Note that the auxiliary variables $\bm{u}$, $\bm{v}$, $\bm{P}$ and $x$ are further defined in Appendix \ref{sec:gradient} for the sake of brevity of the expressions in the table.
\end{tablenotes}
\end{threeparttable}
\end{adjustbox}
%
\vspace{1ex}
{\color{black}\caption{Simulation Parameters}
\vspace{-2ex}
\begin{tabular}{|c|c|c|}
\hline
Reference frames & Translations & Rotations\\
\hline
\setlength{\arraycolsep}{2pt} 
$\bm{C}_1 \!=\!\!
\resizebox{0.39 \textwidth}{!}{$\left[\begin{array}{lllllllc}
-0.5 &\! \phantom{-}0.5 &\! \phantom{-}0.5 &\! -0.5 &\! -0.5 &\! \phantom{-}0.5 &\! -0.5 &\! \phantom{-}0.5 \\
-0.5 &\!-0.5 &\! \phantom{-}0.5 &\!\phantom{-}0.5 &\! -0.5 &\! -0.5 & \!\phantom{-}0.5 & \!\phantom{-}0.5 \\
-0.5 &\! -0.5 &\! -0.5 &\! -0.5 & \!\phantom{-}0.5 &\! \phantom{-}0.5 &\! \phantom{-}0.5 &\! \phantom{-}0.5
\end{array}\right]
$} \!\!\in\! \mathbb{R}^{3 \times 8}\!$ & $\bm{t}_1=[-3,0.5, 7]^\intercal$ &     $[\alpha_1,\beta_1,\gamma_1]=[10^\circ,20^\circ,45^\circ]$
\\
\setlength{\arraycolsep}{2.5pt} 
$\bm{C}_2 \!=\!\!
\resizebox{0.38 \textwidth}{!}{$
\left[\begin{array}{lllllllc}
-10 &\! \phantom{-}10 &\! \phantom{-}10 &\! -10 & \!-10 & \!\phantom{-}10 &\! -10 &\! \phantom{-}10 \\
-10 & \!-10 & \!\phantom{-}10 &\!\phantom{-}10 &\! -10 & \!-10 & \!\phantom{-}10 & \!\phantom{-}10 \\
-10 & \!-10 &\! -10 &\! -10 & \!\phantom{-}10 &\! \phantom{-}10 &\! \phantom{-}10 & \!\phantom{-}10
\end{array}\right]
$}  \!\!\in\! \mathbb{R}^{3 \times 8}$
&
\begin{tabular}{@{}c@{}c@{}}
$\bm{t}_2=[0,0,0]^\intercal$\Tstrut
\vspace{1ex}
\end{tabular}
&
\begin{tabular}{@{}c@{}c@{}}
$[\alpha_2,\beta_2,\gamma_2]=[0^\circ,0^\circ,0^\circ]$
\end{tabular}\\
\hline
\end{tabular}
\label{table::parameters}}
\vspace{-1ex}
\end{table*}
}


{\color{black}
\subsection{CRLB Approximation: Relevance and Complexity Analysis}

To avoid the matrix inversion in the calculation of the \ac{CRLB}, an approximation can be used, as shown in Appendix \ref{sec:AP_Approximation} that leads to the approximated expression for the \ac{CRLB} for the translation vector $\bm{t}$, given by
\begin{equation}
\label{eq:SimpleCRLB_t}
\bar{\omega}_{\bm{t}}\approx \frac{\eta}{\text{tr}(\bm{F}_{\bm{t}})},
\end{equation}
which is shown for completion, but is not worthwhile, since the matrix inversion in equation \eqref{eq:t_CRLB} is not computationally expensive, and thus, the approximation is not needed.

Similar to the case of the translation vector, since the inverse of the \ac{FIM} is computationally expensive to calculate, the same approximation of the \ac{CRLB}, using the relation of the trace of the \ac{FIM} and its inverse can be applied.
Following the same approach as before, it can be shown that after the geometric-arithmetic mean inequality is applied, the following inequality can be found
\begin{equation}
    \frac{\text{tr}(\bm{F}^{-1}_{\bm{Q}})}{\eta^2}\leq\frac{\eta^2}{\text{tr}(\bm{F}_{\bm{Q}})},
\end{equation}
which leads to an average CRLB approximation of
\begin{equation}
\label{eq:SimpleCRLB_Q}
    \color{black}
\bar{\omega}_{\bm{Q}}\approx \frac{\eta^2}{\text{tr}(\bm{F}_{\bm{Q}})}.
\end{equation}
}
\vspace{-3ex}
{\color{black}

We remark that these approximations can be used to simplify the mathematical formulation of algorithms aimed at improving the performance of \ac{RBL} schemes by optimizing system parameters such as anchor locations and the types of measurements that they should collect.
Notice that although techniques to do so in the case of point-target localization exits, $e.g.$, \cite{ShengTAES2017, SaeedTCOM2019, SadeghiTSP2020, Ozturk2021}, these techniques are fundamentally limited to very specific scenarios and cannot be generalized to systems employing multiple types of input information, nor to the \ac{RBL} problems.
In contrast, the information-centric method here contributed and the approximations in equations \eqref{eq:SimpleCRLB_t} and \eqref{eq:SimpleCRLB_Q} leave the possibility for the Frame-theoretical anchor placement optimization approach for point-target localization developed in \cite{Velde_14} to be extended to \ac{RBL} problems, which shall be addressed in a follow-up work.

In terms of computational complexity, the proposed approximation of the \ac{CRLB} requires only the computation of the trace of the \ac{FIM}, which has a computational complexity of $\mathcal{O}(l)$, where $l$ is the dimension of the \ac{FIM}, while the computation of the inverse of the \ac{FIM} in the exact \ac{CRLB} has a computational complexity of $\mathcal{O}(l^3)$, due to the matrix inversion operation.
Thus, the proposed approximation significantly reduces the computational burden, especially for large-scale problems where the dimension of the \ac{FIM} is high.
Notice, however that, as shown in Figures 3 and 4, the performance gap between the exact and approximate solutions decreases as the dimension of the FIM increases.
Consequently, although the approximation is computationally efficient, it does not offer a great advantage over the exact solution in the rigid body localization problem, where the dimension of the \ac{FIM} is relatively small, it is still useful in scenarios where computational resources are limited and the dimension of the \ac{FIM} is larger, which opens up future research directions and optimization problems.

}
\vspace{-1ex}

\section{Numerical Results}
\label{sec:res}

In this section we provide simulation results illustrating the outcome of the proposed \acf{CRLB} for the \ac{RBL} problem, comparing them to \ac{SotA} \ac{RBL} methods.
The bounds are calculated as the {\color{black}} average \ac{CRLB} {\color{black} in order to be represented as the \ac{PEB}}, {\color{black} such that it can be compared to the \ac{RMSE} of the \ac{SotA}} translation vector $\bm{t}$ and the rotation matrix $\bm{Q}$ {\color{black} estimates}.
The used system parameters are described in Table \ref{table::parameters}, unless stated otherwise.

\subsection{Complete and Incomplete Information}
To compare the performance of the various \ac{SotA} methods to the proposed bounds, in terms of translation vector and rotation matrix estimates, we consider the methods proposed in \cite{Fuehrling_2025,Chen_2015} respectively, estimating the parameters by measuring zero mean Normal distributed distance measurements with variance $\sigma^2$.
To be precise, we compare the \ac{CRLB} in terms of the exact representation and the approximation that does not require the inverse of the \ac{FIM}.
Additionally, for the rotation matrix, we also provide the constrained \ac{CRLB} where the rotation matrix is orthogonal and has a determinant of 1, $i.e.$, $\bm{Q}\in\mathcal{SO}(3)$.

\begin{figure}[H]
\centering
{{\includegraphics[width=\columnwidth]{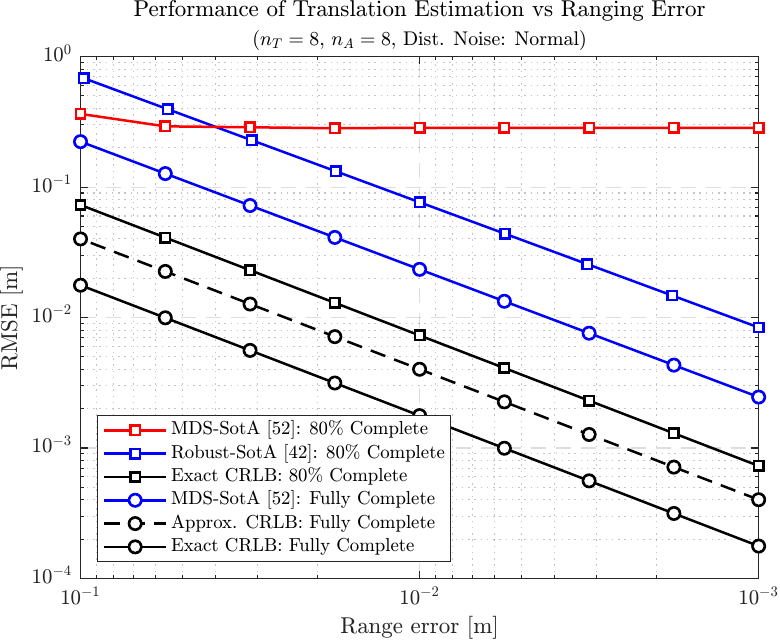}}}
\vspace{-4ex}
\caption{{\color{black}\ac{RMSE}} of the translation vector estimate of the \ac{SotA} and the \ac{CRLB} variations, over the range error $\sigma$.}
\label{fig:Tra}
\end{figure}

\begin{figure}[H]
{{\includegraphics[width=\columnwidth]{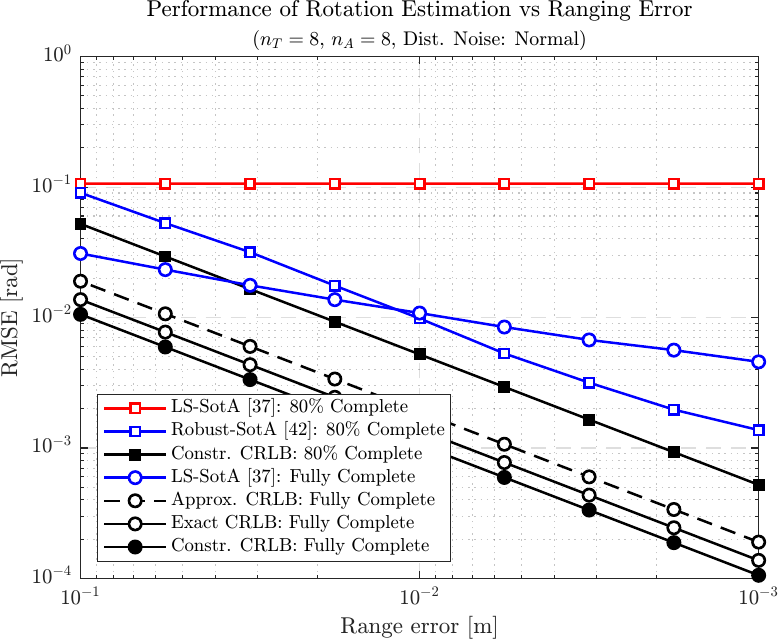}}}
\vspace{-4ex}
\caption{{\color{black}\ac{RMSE}} of the rotation matrix estimate of the \ac{SotA} and the \ac{CRLB} variations, over the range error $\sigma$.}
\label{fig:Rot}
\end{figure}

The first set of results is shown in Figure \ref{fig:Tra} and Figure \ref{fig:Rot}, illustrating the performance of conventional distance-based \ac{RBL} methods for complete and incomplete information, compared to the corresponding bounds.
In Figure \ref{fig:Tra}, the \ac{CRLB} for the translation vector $\bm{t}$ is shown as a function of the range error $\sigma$ for the exact \ac{CRLB}, the approximation of it, the \ac{MDS} based method proposed in \cite{Nic_RBL} as well as the bound for incomplete measurements, where only 80\% of the possible measurements were observed, compared to the incomplete \ac{MDS} based method, as well as a robust \ac{RBL} method proposed in \cite{Fuehrling_2025}.
It can be observed that if all measurements are available the \ac{MDS}-\ac{SotA} performs well, but not close to the corresponding bound.
While the incomplete version of the \ac{MDS}-\ac{SotA} converges to a poor error floor, the robust method only performs slightly worse than the fully connected scenario, even when only 80\% of the observations are available, still not very close to the bound, but closer than in the fully connected case.

Next, in Figure \ref{fig:Rot}, the \ac{CRLB} for the rotation matrix $\bm{Q}$ is shown as a function of the range error $\sigma$ for the exact unconstrained \ac{CRLB}, the unconstrained approximation and the constrained \ac{CRLB}, compared to the least squares solution of \cite{Chen_2015}, as well as the scenario of incomplete observations, again comparing the bound to the robust method of \cite{Fuehrling_2025}.
It can be observed that the approximation of the unconstrained bound is very close to the exact bound, while the constrained \ac{CRLB} is slightly lower than the unconstrained {\color{black}bound}, which is expected since the constrained \ac{CRLB} makes use of the orthogonality condition of the rotation matrix, which {\color{black}leads} to a lower bound.

\vspace{-1ex}

For the fully connected scenario, compared to the actual estimate of the rotation matrix, especially in low error regimes, the \ac{CRLB} is significantly lower, which hints at the potential that is left to improve the performance of the \ac{RBL} methods, as well as the fact that only leveraging distance measurements is not optimal for rotation estimation.
However, in the case of incomplete observations with 80\% observed measurements, while the least square solution performs poorly at a high error floor, the robust method performs close to the corresponding bound for all error ranges, while outperforming the least square solution for lower error regimes.

All in all, it can be observed that for both, the translation and rotation, the estimates of most \ac{SotA} methods are rather poor compared to the bound and that the bound degrades due to incomplete information, which indicates that there is still a lot of room for improvement, and the performance of the \ac{RBL} methods can be improved significantly\footnotemark.
Nevertheless, as shown above, recent contributions already deal with scenarios of incomplete observations and can already get close to the \ac{CRLB} for the rotation matrix estimate.

\vspace{-2ex}

\subsection{Evaluation with Heterogeneous Information}

\footnotetext{While the performance of the \ac{RBL} methods is closer to the \ac{CRLB} for high range errors, the \ac{CRLB} is not necessarily a good indicator of achievable accuracy in this regime due to its reliance on small-errors, which is the reason why we do not show the results for high range errors.}

To showcase that the proposed \ac{CRLB} framework is not limited to distance measurements, but can also be adopted to other types of measurements and specifically heterogeneous information, in this section we consider a \ac{RBL} scenario where the measurements are a combination of distance and angle measurements, such that the rigid bod can be estimated by an \ac{SMDS}-based approach{\color{black}, presented in \cite{Nic_ICNC}}, calculating the bound with respect to the observed measurements between the nodes.

\begin{figure}[H]
\centering
\includegraphics[width=0.8\columnwidth]{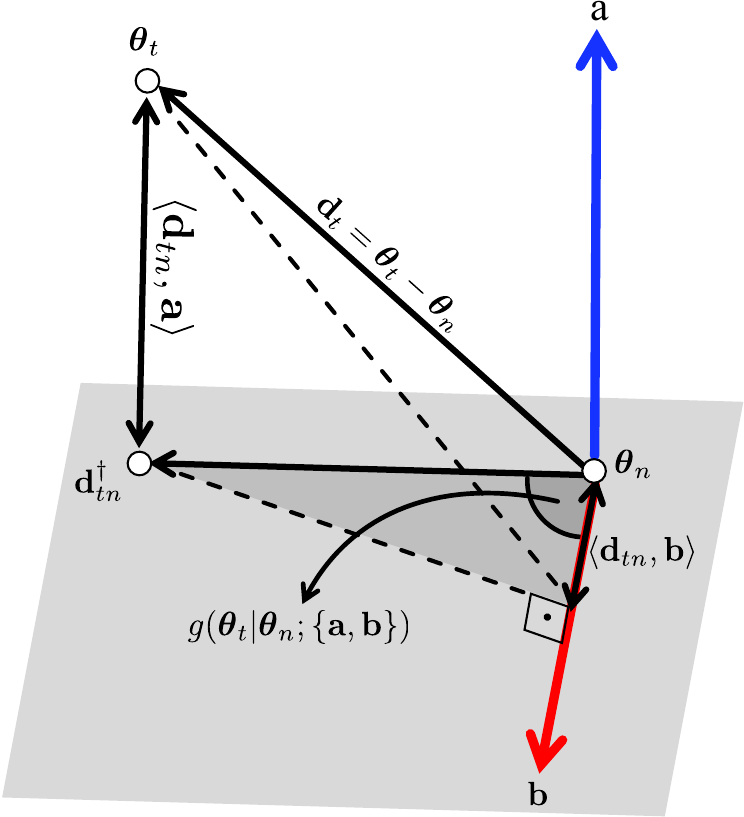}
\caption{Illustration of the angle-of-arrival measurements dissimilarity function calculation. Notice that the orthonormal vectors $\mathbf{a}$ and $\mathbf{b}$ are centered at $\bm{\theta}_a$, and act as a base for the angle of arrival measurement. Specifically, the plane onto which $\mathbf{d}_{na}$ is projected (giving $\mathbf{d}^\dagger_{na}$) is defined by (normal to) $\mathbf{a}$ and contains the reference vector $\mathbf{b}$ against which the angle of arrival is measured.}
\label{Fig:IllustrationAngleDissimilarity}
\end{figure}

To that extend, consider a scenario, where the nodes perform not only distance measurements, but also \ac{AoA} measurements, as shown in Figure \ref{Fig:IllustrationAngleDissimilarity}.
We therefore consider Gamma distributed distance measurements and Von-Mises distributed angle measurements, which illustrates the applicability of the bound to different types of measurements with different error distributions.

\begin{figure}[H]
\centering
{{\includegraphics[width=\columnwidth]{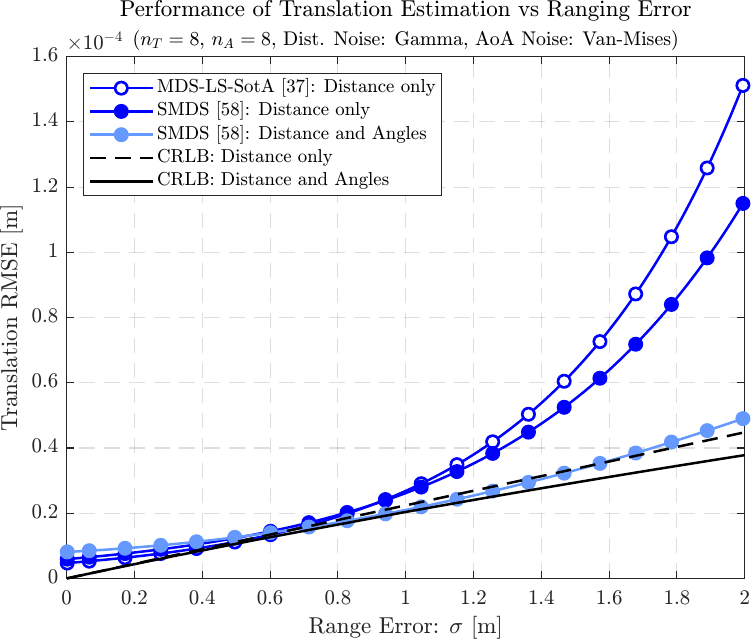}}}
\vspace{-4ex}
\caption{{\color{black}\ac{RMSE}} of the translation vector estimate of the \ac{SotA} and the \ac{CRLB} variations, over various noise levels.}
\label{fig:t_Hetero}
\end{figure}

\begin{figure}[H]
\centering
{{\includegraphics[width=\columnwidth]{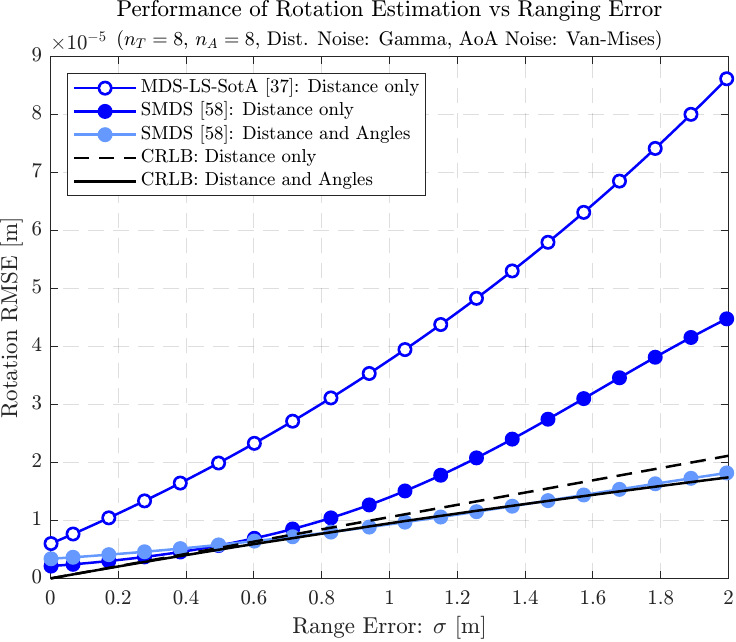}}}
\vspace{-4ex}
\caption{{\color{black}\ac{RMSE}} of the rotation matrix estimate of the \ac{SotA} and the \ac{CRLB} variations, over various noise levels.}
\label{fig:Rot_Hetero}
\vspace{-1ex}
\end{figure}

Figure \ref{fig:t_Hetero} and \ref{fig:Rot_Hetero} show the result of the translation and rotation estimation for an \ac{MDS}-based approach, an \ac{SMDS}-based approach, where the angles are obtained by using only the distance measurements, and a full \ac{SMDS} approach, where both distance and angle measurements are used, compared to the corresponding \ac{CRLB}, only with distance measurements and with both distance and angle measurements.  
It can be observed that the distance only \ac{SMDS} approach is performing better than the \ac{MDS} approach, which is expected since the \ac{SMDS} approach uses more information, artificially created by estimating the angles, and thus, should lead to a better estimate.
However, the distance only \ac{SMDS} approach is not performing as well as the full \ac{SMDS} approach, which is also expected, since the full \ac{SMDS} approach uses both distance and angle measurements.


In Figure \ref{fig:t_Hetero} it can be observed that in general, the full \ac{SMDS} yields the best results with estimates close to the \ac{CRLB}, while the \ac{SMDS} approach with only distance measurements is performing better than the \ac{MDS} approach, and slightly better that the full \ac{SMDS} in small range error regimes, which is expected, since in \cite{Abreu_2007} it was shown that the \ac{SMDS} approach is not optimal in the small range error regime, and thus, the \ac{MDS} approach performs better in this regime.
Nevertheless, the distance only \ac{SMDS} approach performs close to the \ac{CRLB} in the small range error regime, which indicates that the \ac{SMDS} approach is a good choice for the rigid body localization problem, even with only distance measurements.

\vspace{-1ex}

In Figure \ref{fig:Rot_Hetero} similar properties can be observed, where the full \ac{SMDS} approach is performing best, while the \ac{SMDS} approach with only distance measurements is performing much better than the \ac{MDS} approach, with the full \ac{SMDS} approach is very close to the \ac{CRLB} over the whole error range, while the \ac{SMDS} approach with only distance measurements also performs close to the \ac{CRLB} in the small range error regimes.

\vspace{-2ex}
\section{Conclusion}
\label{sec:conclusions}
We proposed a novel information-centric framework for the construction of the \ac{CRLB} for \ac{RBL}, leading to a generalized \ac{CRLB} for the rigid body parameters that includes the rotation matrix and translation vector applicable to any type of measurements.
To that extend, first the classical element-centric \ac{FIM} was presented and extended to the information-centric construction approach, leading to a generalized \ac{FIM} framework {\color{black}that is applicable to arbitrary types of information and measurement error distributions and enables a straightforward adjustment of the bound when measurements are added or removed}.
Then the \ac{FIM} was applied to the \ac{RBL} problem, by making use of the information-gradient and information-intensity concepts, by exploiting the Frobenius inner product and the differential approach for derivatives, which lead to a generalized \ac{CRLB} for the rigid body parameters. 
Furthermore, in addition to an approximate solution for the \ac{CRLB} that does not require the inverse of the \ac{FIM}, we also derived a constrained \ac{CRLB} that ensures that the rotation matrix is orthogonal and has a determinant of 1, $i.e.$, $\bm{Q}\in\mathcal{SO}(3)$.
Simulation results illustrate the great potential of \ac{RBL} by using the proposed \ac{CRLB} framework, showing that the proposed \ac{CRLB} is a suitable bound for the \ac{RBL} problem, not only in the case of distance measurements, but also in the case of heterogeneous information, such as a mixture of distance and angle-of-arrival measurements.
However, comparing the bound to prior \ac{SotA} methods, we show that the proposed \ac{CRLB} is not tight in the low range error regimes for all \ac{SotA} methods, and thus, further work on \ac{RBL} estimators is needed to improve the performance of the estimation algorithms.


\vspace{-2ex}
\appendices
\section{Fisher Information Derivations}
\vspace{-1ex}

As shown in Table \ref{TAB_InformationIntensities}, the information intensity can be derived for different types of error statistics.
As examples, in this section we derive the information intensities for  Von-Mises \cite{Abreu2008Tikhonov}, Nakagami \cite{NakagamiFunc_Moment12}, and Gamma variables, which can be directly plugged into the \ac{FIM} construction framework.

\vspace{-2ex}
\subsection{Von Mises Variables}
\label{SEC_VonMisesFisher}
\vspace{-1ex}

Let us start with the Von-Mises distribution (Tikhonov
distribution) \cite{Abreu2008Tikhonov}, which can be defined as
\begin{equation}
p(r_{na};g(\bm{\theta}_n\vert \bm{\theta}_a)) = \dfrac{1}{2\pi I_{0}(\omega)} e^{\omega\,\mathrm{cos}(r_{na}- \mu)},
\end{equation}
where the variation function lies in the range of$-\pi\leq r_{na}\leq\pi$, while $\omega$ and $\mu$ are the shape and centrality parameters of the distribution and $I_0(x)$ denotes the modified Bessel function of the first kind and $0$-th order \cite{Bowman_2012}.

From the above distribution, the corresponding Fisher information can be found according to equation \eqref{eq:Fisher_Info}, yielding
\begin{eqnarray}
F = \mathbb{E}_{r_{na}}\!\!\left[\left(\frac{\partial \ln p(r_{na};g(\bm{\theta}_n\vert \bm{\theta}_a))}{\partial g(\bm{\theta}_n\vert \bm{\theta}_a)}\right)^{2}\right]&&\nonumber \\
&&\hspace{-40ex} = \mathbb{E}_{r_{na}}\!\!\left[\left(-\omega\sin\left(r_{na}\!-\!\mu \right)\right)^{2}\right]
= \omega^{2}\mathbb{E}_{r_{na}}\!\Big[\frac{1\!-\!\cos\left(2(r_{na}\!-\!\mu)\right)}{2}\Big]\nonumber \\
&&\hspace{-40ex}= \frac{\omega^{2}}{2} - \int\limits_{-\pi}^{\pi} \frac{\cos\left(2(r_{na}-\mu)\right)}{2\pi I_{0}(\omega)} e^{\omega \cos(r_{na}-\mu)} \, dr_{na}
= \frac{\omega^{2}}{2},
\end{eqnarray}
where we made use the geometric identity $\mathrm{sin}^{2}(x)=\frac{1-\mathrm{cos}(2x)}{2}$
and the fact that $\mathrm{cos}(2x)$ and $\mathrm{exp}\left(\omega\mathrm{cos}(x)\right)$
are two orthogonal functions for which the integral over $[0,\,2\pi]$ vanishes, which leads to an information intensity of
\begin{equation}
\lambda=\sqrt{F}=\frac{\omega}{\sqrt{2}}.
\end{equation}

\vspace{-3ex}
\subsection{Nakagami Variables}
\label{SEC_NakagamiFisher}
\vspace{-1ex}

Next, the distribution of the Nakagami-$m$ variable \cite{NakagamiFunc_Moment12} can be investigated, where the corresponding \ac{PDF} is defined as
\begin{equation}
p(r_{na};g(\bm{\theta}_n\vert \bm{\theta}_a)) = \dfrac{2m^{m}}{\Gamma(m) \Upsilon^{m}} r_{na}^{2m-1} \exp{\left(-\dfrac{m}{\Upsilon}r_{na}^2\right)},
\end{equation}
where $m = \Upsilon^2 / \mathbb{E}\left[ (r_{na}^2-\Upsilon)^2 \right]  $ is the shape parameter with $m \geq \frac{1}{2}$, $\Upsilon = \mathbb{E} \left[r_{na}^2\right] \geq 0$ is the second moment of the distribution that controls the spread and  $\Gamma(.)$ is the Gamma function \cite{abramowitz1965handbook}.

Similarly to before, we follow equation \eqref{eq:Fisher_Info}, to define the \ac{FIM} corresponding to the Nakagami-$m$ distribution as
\begin{equation}
\begin{split}
F &= \mathbb{E}_{r_{na}}\left[\left(\frac{\partial \ln p(r_{na};g(\bm{\theta}_n \vert \bm{\theta}_a))}{\partial g(\bm{\theta}_n \vert \bm{\theta}_a)}\right)^{2}\right] \\
&= \mathbb{E}_{r_{na}}\left[\left(\frac{2m - 1}{r_{na}} - \frac{2m r_{na}}{\Upsilon} \right)^2 \right] \\
&= (2m - 1)^2 \, \mathbb{E}_{r_{na}}\!\left[r_{na}^{-2}\right]
+ \frac{4m^2}{\Upsilon^2} \, \mathbb{E}_{r_{na}}\!\left[r_{na}^2\right]
- \frac{8m^2 - 4m}{\Upsilon} \\
&= (2m - 1)^2 \cdot \frac{m}{(m - 1)\Upsilon}
+ \frac{4m^2}{\Upsilon^2} \cdot \Upsilon
- \frac{8m^2 - 4m}{\Upsilon} \\
&= \frac{m(4m - 3)}{\Upsilon(m - 1)},
\end{split}
\end{equation}
which leads to an information intensity of
\begin{equation}
    \color{black}
\lambda=\sqrt{F}=\sqrt{\frac{m(4m - 3)}{\Upsilon(m - 1)}}.
\end{equation}

In the above proof, to derive the first term of the expression, we used the general integral property \cite[p361]{IntegralsJeffrey}
\begin{equation}
\label{EQ_integral_Property}
\int_{0}^{\infty}x^{\alpha-1}e^{-px^\mu} dx = \frac{1}{\mu}p^{-\alpha/\mu}\Gamma\left(\frac{\alpha}{\mu}\right),
\end{equation}
while for the second term we employed the formula for the $k$-th moment of Nakagami distribution {\color{black}$\mathbb{E}[r_{na}^k]=\mleft(\frac{\Upsilon}{m}\mright)^{\frac{k}{2}}\frac{\Gamma(m+k/2)}{\Gamma(m)}$}  \cite{NakagamiFunc_Moment12}.
\vspace{-1ex}

\subsection{Gamma Variables}
\label{SEC_GammaFisher}
\vspace{-1ex}

Finally, we can investigate information intensity of the Gamma distribution.
Although there exists multiple parameterization of the Gamma distribution, we choose to use the one with the corresponding PDF given by
\begin{equation}
p(r_{na};g(\bm{\theta}_n\vert \bm{\theta}_a)) = \dfrac{1}{\Gamma(\kappa) \upsilon^{\kappa}} r_{na}^{\kappa-1} \exp{\left(-\dfrac{r_{na}}{\upsilon}^2\right)},
\end{equation}
where $\kappa>0$ and $\upsilon>0$ are the shape and scale parameter accordingly. Similar to the above sections and following equation \eqref{eq:Fisher_Info}, the \ac{FIM} corresponding to Gamma distribution can be derived as
\begin{equation}
\begin{split}
F &= \mathbb{E}_{r_{na}}\left[\left(\frac{\partial \ln p(r_{na};g(\bm{\theta}_n\vert \bm{\theta}_a))}{\partial g(\bm{\theta}_n\vert \bm{\theta}_a)}\right)^{2}\right] \\
&= \mathbb{E}_{r_{na}}\left[\left(\frac{\kappa-1}{r_{na}} - \frac{1}{\upsilon}\right)^2 \right] \\
&= (\kappa - 1)^2 \, \mathbb{E}_{r_{na}}\!\left[r_{na}^{-2}\right]
- \frac{2(\kappa - 1)}{\upsilon} \, \mathbb{E}_{r_{na}}\!\left[r_{na}^{-1}\right]
+ \frac{1}{\upsilon^2} \\
&= (\kappa - 1)^2 \cdot \frac{1}{\upsilon^2(\kappa - 1)(\kappa - 2)}
- \frac{2(\kappa - 1)}{\upsilon} \cdot \frac{1}{\upsilon(\kappa - 1)}
+ \frac{1}{\upsilon^2} \\
&= \frac{1}{\upsilon^2(\kappa - 2)},
\end{split}
\end{equation}
\vspace{-3ex}
with the information intensity given by
\begin{equation}
\lambda=\sqrt{F}=\sqrt{\frac{1}{\upsilon^2(\kappa-2)}}.
\end{equation}

Again, in the above proof, to derive the first and second term of the expression, we used the general integral property from equation \eqref{EQ_integral_Property}. 
\vspace{-5ex}

\section{Gradient derivation for Rigid Body Parameters}
\label{sec:gradient}
\vspace{-1ex}

To derive the gradient of the dissimilarity function with respect to the rigid body parameters, we consider the differential approach {\color{black}for matrix calculus \cite{Magnus_2019}}, which allows us to compute the gradient of the dissimilarity function for different types of measurements, by using the properties of the double dot product.
The double dot product, also known as the Frobenius inner product{\color{black}\cite{Herzog_2025}}, is a way to compute the inner product of two matrices by treating them as vectors in a higher-dimensional space.
For two \( m \times n \) matrices \( \mathbf{A} \) and \( \mathbf{B} \), the Frobenius inner product is defined as
\begin{equation}
    \color{black}
     \mathbf{A}:\mathbf{B}  = \sum_{i=1}^m \sum_{j=1}^n \mathbf{A}_{ij} \mathbf{B}_{ij} = \text{tr}(\mathbf{A}^{\color{black}\intercal} \mathbf{B}),
\end{equation}
where  $\mathbf{A}_{ij}$ and $\mathbf{B}_{ij}$ are the elements of $\mathbf{A}$ and $\mathbf{B}$, and $\text{tr}$ denotes the trace (sum of diagonal elements).
This is equivalent to the sum of the element-wise products or the trace of the matrix product ${\color{black}\mathbf{A}^{\color{black}\intercal} \mathbf{B}}$.

Similar to multiplication properties, the Frobenius inner product satisfies the commutative property $\mathbf{A} : \mathbf{B} = \mathbf{B} : \mathbf{A}$, the distributive property $\langle a\mathbf{A} + b\mathbf{C}, \mathbf{B} \rangle_F = a \langle \mathbf{A}, \mathbf{B} \rangle_F + b \langle \mathbf{C}, \mathbf{B} \rangle_F$, for scalars $a, b$ and matrices $\mathbf{A}, \mathbf{C}, \mathbf{B}$, and specifically the following property to rearrange the matrices in the inner product $\mathbf{C}\mathbf{A}:\mathbf{B}=\mathbf{C}:\mathbf{B}\mathbf{A}^\intercal=\mathbf{A}:\mathbf{C}^\intercal \mathbf{B}$.

\subsection{{\color{black}Square} Distance Measurements}
\vspace{-1ex}

To begin, consider the dissimilarity function $ g(\bm{\theta}_n\vert \bm{\theta}_a) $ defined as the squared Euclidean distance between two points in a rigid body localization scenario, where $ \bm{\theta}_n $ is the target position and $ \bm{\theta}_a $ is the anchor position, given by
\begin{equation}
    g(\bm{\theta}_n\vert \bm{\theta}_a)=\|\bm{\theta}_n-\bm{\theta}_a\|^2.
\end{equation}

To simplify the notation, let us denote an auxiliary variable $\bm{u} = \bm{\theta}_n-\bm{\theta}_a = \bm{Q}\bm{c}_t + \bm{t} - \bm{\theta}_a $, such that the dissimilarity function can be rewritten as
\begin{equation}
    \color{black}
    g = \bm{u}^{\color{black}\intercal} \bm{u} = \bm{u}:\bm{u},
\end{equation}
where $ \bm{u}:\bm{u} $ denotes the double dot product of the vector $ \bm{u} $ with itself.

Next, the differential of the dissimilarity function with respect to $\bm{u}$ can be computed as
\begin{equation}
    dg = 2 \bm{u}:\, d\bm{u}.
\end{equation}

To obtain the gradient of the dissimilarity function with respect to the parameters of interest, we need to compute the differential $ d\bm{u} $ for the two different cases, $i.e.$, translation and rotation.
In the first case, we consider the translation of the target position, where the differential can be expressed as
\begin{equation}
    d\bm{u} = d\bm{t}.
\end{equation}
Following the rules of using the double dot product for gradients using differentials, we can insert the differential into the expression for the dissimilarity function, leading to
\begin{equation}
    dg = 2 \bm{u}:\, d\bm{t},
\end{equation}
which leads to the gradient of the dissimilarity function with respect to the translation vector $\bm{t}$ as
\begin{equation}
    g' \big|_{\mathbf{t}}^{d}=\frac{\partial g(\bm{\theta}_n\vert \bm{\theta}_a)}{\partial \bm{t}} = 2 \bm{u} = 2 (\bm{Q}\bm{c}_t + \bm{t} - \bm{\theta}_a).
\end{equation}

In the second case, we consider the vectorized rotation matrix $\text{vec}(\bm{Q})$, where first, the auxiliary variable $\bm{u}$ can be rewritten as
\begin{equation}
\bm{u}=(\bm{c}_t^\intercal \otimes \bm{I}_{\eta}) \text{vec}(\bm{Q}) + \bm{t} - \bm{\theta}_a,
\end{equation}
with the corresponding differential given by
\begin{equation}
    d\bm{u} = (\bm{c}_t^\intercal \otimes \bm{I}_{\eta}) d\text{vec}(\bm{Q}).
\end{equation}

Finally, inserting the differential into the expression for the dissimilarity function, we obtain
\begin{equation}
    dg = 2 \bm{u}:\, (\bm{c}_t^\intercal \otimes \bm{I}_{\eta}) d\text{vec}(\bm{Q}),
\end{equation}
which leads to the gradient of the dissimilarity function with respect to the vectorized rotation matrix $\text{vec}(\bm{Q})$ as
\begin{equation}
    g' \big|_{\mathbf{Q}}^{d}=\frac{\partial g(\bm{\theta}_n\vert \bm{\theta}_a)}{\partial \text{vec}(\bm{Q})} = 2 (\bm{c}_t \otimes \bm{I}_{\eta}) (\bm{Q}\bm{c}_t + \bm{t} - \bm{\theta}_a).
\end{equation}

\subsection{Angle of arrival measurements}
\vspace{-1ex}

Next, the dissimilarity function for the \acf{AoA} measurements is considered, which is defined as the angle between two vectors in a rigid body localization scenario, as illustrated in Figure \ref{Fig:IllustrationAngleDissimilarity}.
The dissimilarity function is given by$ g(\boldsymbol{\theta_t}| \boldsymbol{\theta}_n, \boldsymbol{a}, \boldsymbol{b}) $ is defined as
\begin{equation}
g(\boldsymbol{\theta}_t| \boldsymbol{\theta}_n, \boldsymbol{a}, \boldsymbol{b}) = \arccos\left( \frac{\langle \boldsymbol{d}_{na}, \boldsymbol{b} \rangle}{\|\boldsymbol{d}_{na}- \langle \boldsymbol{d}_{na},\boldsymbol{a} \rangle\boldsymbol{a}\|} \right),
\end{equation}
where again, an auxiliary variable can be defined as
\begin{equation}
\bm{u}=\boldsymbol{d}_{na}= \bm{\theta}_n-\bm{\theta}_a = \bm{Q}\bm{c}_t + \bm{t} - \bm{\theta}_a,
\end{equation}
such that the dissimilarity function can be rewritten as
\begin{equation}
g = \arccos\left( \frac{\bm{u}^\intercal\boldsymbol{b}}{\|\bm{u} - \bm{u}^\intercal\boldsymbol{a}\boldsymbol{a}\|} \right)=\arccos(x).
\end{equation}

To avoid the chain rule, the differential approach enables the step by step computation of the differentials until the desired parameter is reached, such that the final gradient can be computed.
Thus, to start with, the differential of the dissimilarity function with respect to $ x $ can be computed as
\begin{equation}
    \label{eq:diff_arccos}
{d g} = -\frac{1}{\sqrt{1 - x^2}} {d x},
\end{equation}
where the variable $ x $ is defined as
\begin{equation}
x = \frac{\bm{u}^\intercal\boldsymbol{b}}{\|\bm{u} - \bm{u}^\intercal\boldsymbol{a}\boldsymbol{a}\|} = \frac{\boldsymbol{b}}{\|\bm{u} - \bm{u}^\intercal\boldsymbol{a}\boldsymbol{a}\|} : \bm{u} = \frac{\boldsymbol{b}}{\|(\bm{I}-\boldsymbol{a}\boldsymbol{a}^\intercal)\bm{u}\|} : \bm{u}.
\end{equation}  
Next, the differential of $ x $ with respect to $ \bm{u} $ can be computed, where the projection matrix $ \bm{P}= \bm{I}-\boldsymbol{a}\boldsymbol{a}^\intercal $ is used to simplify the expression, leading to
\begin{equation}
    \color{black}
d x = \frac{(\bm{u}^{\color{black}\intercal} \bm{P} \bm{u} ) \boldsymbol{b} - ( \bm{u}^{\color{black}\intercal} \boldsymbol{b} )\cdot \bm{P} \bm{u}}{(\|\bm{P} \bm{u}\|)^3}: d \bm{u}.
\end{equation}

Inserting the differential into the expression for the dissimilarity function, we obtain
\begin{equation}
    \color{black}
{d g} = -\frac{1}{\sqrt{1 - x^2}} \cdot \underbrace{\left(\frac{(\bm{u}^{\color{black}\intercal} \bm{P} \bm{u} ) \boldsymbol{b} - ( \bm{u}^{\color{black}\intercal} \boldsymbol{b} )\cdot \bm{P} \bm{u}}{(\|\bm{P} \bm{u}\|)^3}\right)}_{\bm{w}\in\mathbb{R}^{3\times1}} : d \bm{u},
\end{equation}
where again, the differential $ d \bm{u} $ can be computed for the two different cases, $i.e.$, translation and rotation, as before.
Thus, for the translation case, the differential expression for the dissimilarity function as
\begin{equation}
{d g} = -\frac{1}{\sqrt{1 - x^2}} \cdot \bm{w} : d \bm{t},
\end{equation}
which leads to the gradient of the dissimilarity function with respect to the translation vector $ \bm{t} $ as
\begin{equation}
    \label{eq:grad_AoA_translation}
g' \big|_{\mathbf{t}}^{\psi}=\frac{\partial g(\boldsymbol{\theta_t}| \boldsymbol{\theta}_n, \boldsymbol{a}, \boldsymbol{b})}{\partial \bm{t}} = -\frac{1}{\sqrt{1 - x^2}} \cdot \bm{w} \in \mathbb{R}^{3\times1}.
\end{equation}

In the second case, for the rotation case, the differential expression for the dissimilarity function as \vspace{-1ex}
\begin{equation}            
{d g} = -\frac{1}{\sqrt{1 - x^2}} \cdot \bm{w} : (\bm{c}_t^\intercal \otimes \bm{I}_{\eta}) d \text{vec}(\bm{Q}),
\end{equation}
which leads to the gradient of the dissimilarity function with respect to the vectorized rotation matrix $ \text{vec}(\bm{Q})$ as\vspace{-1ex}
\begin{equation}
    \label{eq:grad_AoA_rotation}
g' \big|_{\mathbf{Q}}^{\psi}=\frac{\partial g(\boldsymbol{\theta_t}| \boldsymbol{\theta}_n, \boldsymbol{a}, \boldsymbol{b})}{\partial \text{vec}(\bm{Q})} = -\frac{(\bm{c}_t \otimes \bm{I}_{\eta})}{\sqrt{1 - x^2}} \cdot \bm{w} \in \mathbb{R}^{9\times 1}.
\end{equation}

\vspace{-2ex}

\subsection{Angle difference of arrival measurements}
\vspace{-1ex}

Finally, the dissimilarity function for the \ac{ADoA} measurements $g(\bm{\theta}_n| {\bm{\theta}}_n,  {\bm{\theta}}_k)$ is considered, given by
\begin{equation}
g(\bm{\theta}_n| {\bm{\theta}}_n,  {\bm{\theta}}_k) = \arccos\left( \frac{d_{na}^2+ d_{ka}^2- d_{nk}^2}{2 d_{na} d_{ka}} \right),
\end{equation}
where the distances are defined as \vspace{-1ex}
\begin{subequations}
\begin{equation}
d_{na}=||\bm{\theta}_n-\bm{\theta}_a||, 
\end{equation}
\begin{equation}
d_{nk}=||\bm{\theta}_n-\bm{\theta}_k||,
\end{equation}
\begin{equation}
d_{ka}=||\bm{\theta}_k-\bm{\theta}_a||,
\end{equation}
\end{subequations}
such that the auxiliary variables can be defined as\vspace{-1ex}
\begin{subequations}
\begin{equation}
\bm{u}= \bm{\theta}_n-\bm{\theta}_a = \bm{Q}\bm{c}_t + \bm{t} - \bm{\theta}_a,
\end{equation}
\begin{equation}
\bm{v}= \bm{\theta}_n-\bm{\theta}_k = \bm{Q}\bm{c}_t + \bm{t} - \bm{\theta}_k,
\end{equation}
\begin{equation}
    d=||\bm{\theta}_k-\bm{\theta}_a||=d_{ka}.
\end{equation}
\end{subequations}

Then, the dissimilarity function can be rewritten as
\begin{equation}
g = \arccos\left( \frac{\bm{u}^\intercal \bm{u}+d^2+\bm{v}^\intercal \bm{v}}{2\|\bm{u}\|d} \right)=\arccos(x).
\end{equation}  

Since the dissimilarity function outer function, the arccos is identical to before, differential of with respect to $ x $ is also identical to equation \eqref{eq:diff_arccos}.
Next, the differential of $ x $ with respect to $ \bm{u} $ can be computed, where first, the variable $ x $  can be rewritten as
\begin{equation}
x = \frac{\bm{u}^\intercal \bm{u}+d^2+\bm{v}^\intercal \bm{v}}{2\|\bm{u}\|d} = \frac{\bm{u}^\intercal \bm{u}}{2\|\bm{u}\|d} + \frac{d^2}{2\|\bm{u}\|d} + \frac{\bm{v}^\intercal \bm{v}}{2\|\bm{u}\|d},
\end{equation}  
such that the differential of $ x $ with respect to $ \bm{u} $ can be computed as\vspace{-1ex}
\begin{equation}
    \label{eq:ADoA_dx}
d x = \frac{\bm{u}}{2\|\bm{u}\|d} :d\bm{u}-\frac{\bm{u}d}{2\|\bm{u}\|^3} :d\bm{u}+\frac{||\bm{v}||^2u}{2\|\bm{u}\|^3d}:d\bm{u}-\frac{\bm{v}}{\|\bm{u}\|d}:d\bm{v},
\end{equation}
where in the last term the chain rule is used to compute the differential with respect to both $ \bm{u} $ and $ \bm{v} $, since both auxiliary variables are dependent on the rigid body parameters.
Nevertheless, the differential of these variables with respect to the rigid body parameters are identical, $i.e.$ $d\bm{u}=d\bm{v}$, such that equation \eqref{eq:ADoA_dx} can be rewritten as \vspace{-1ex}
\begin{equation}
d x = \left(\frac{\bm{u}}{2\|\bm{u}\|d} -\frac{\bm{u}d}{2\|\bm{u}\|^3} +\frac{||\bm{v}||^2u}{2\|\bm{u}\|^3d}-\frac{\bm{v}}{\|\bm{u}\|d}\right):d\bm{u}.
\end{equation}

Next the differential of the dissimilarity function with respect to $ \bm{u} $ can be computed by inserting the differential of $ x $ into equation \eqref{eq:diff_arccos}, leading to
\begin{equation}
{d g} = \frac{-1}{\sqrt{1 - x^2}} \cdot \underbrace{\left(\frac{\bm{u}}{2\|\bm{u}\|d} -\frac{\bm{u}d}{2\|\bm{u}\|^3} +\frac{||\bm{v}||^2u}{2\|\bm{u}\|^3d}-\frac{\bm{v}}{\|\bm{u}\|d}\right)}_{w\in\mathbb{R}^{3\times1}}\!:\! d \bm{u}.
\end{equation}

The final expression for the gradients are identical to the ones shown in equations \eqref{eq:grad_AoA_translation} and \eqref{eq:grad_AoA_rotation}, inserting the variables $x$ and $\bm{w}$ into the expression, leading to $g' \big|_{\mathbf{t}}^{\psi'}$ and $g' \big|_{\mathbf{Q}}^{\psi'}$ respectively.
\vspace{-1ex}

\section{Low-complexity Approximation of CRLB}
\label{sec:AP_Approximation}
\vspace{-1ex}

As mentioned before, since the inverse of the \ac{FIM} is computationally expensive to calculate, we propose an approximation to the \ac{CRLB} by using the relation of the trace of the \ac{FIM} and its inverse, as already used in earlier \ac{SotA} methods.
Thus, consider the following pair of equations for the trace of the \ac{FIM} and its inverse with respect to its eigenvalues $\Upsilon_i$, such that

\begin{subequations}
\begin{equation}
\label{eq:eig_F}
\text{tr}(\bm{F})=\sum_{i=1}^{\eta}\Upsilon_i\triangleq T,
\end{equation}
\begin{equation}
    \label{eq:eig_F_inv}
    \text{tr}(\bm{F}^{-1})=\sum_{i=1}^{\eta}\frac{1}{\Upsilon_i}.
\end{equation}
\end{subequations}

Rewriting equations \eqref{eq:eig_F} and \eqref{eq:eig_F_inv} it can be obtained that
\begin{equation}
    \frac{1}{T} \sum_{i=1}^{\eta } \Upsilon_i=1 \rightarrow \sum_{i=1}^{\eta} \frac{\Upsilon_i}{T}=1 \rightarrow \sum_{i=1}^{\eta} \Upsilon_i^{\prime}=1
\end{equation}

Next, by using the geometric-arithmetic mean inequality, the following inequality can be found
\begin{equation}
    \sum_{i=1}^{\eta } \frac{1}{\Upsilon_i^{\prime}} \geq\eta \rightarrow \sum_{i=1}^{\eta} \frac{1}{\Upsilon_i} \geq \frac{\eta}{T},
\end{equation}
such that after combining the equations, the final relation can be found
\begin{equation}
    \frac{\text{tr}(\bm{F}^{-1})}{\eta}\leq\frac{\eta}{\text{tr}(\bm{F})},
\end{equation}
which leads to an average CRLB approximation of
\begin{equation}
\bar{\omega}\approx \frac{\eta}{\text{tr}(\bm{F})},
\end{equation}
used to approximate the average CRLB for any parameter of interest, without the need to calculate the inverse of the \ac{FIM}, where the nominator depends on the dimension of the \ac{FIM}.





\end{document}